# Combined evanescent-wave excitation and supercritical-angle fluorescence detection improves optical sectioning

Maia Brunstein[a,b,c], Maxime Teremetz[a,b,c,d,1], Christophe Tourain[a,b,c,e], Martin Oheim[a,b,c,2]

[a]*CNRS, UMR 8154, Paris, F-75006 France;* [b]*INSERM, U603, Paris, F-75006 France;*
[c]*Laboratoire de Neurophysiologie et Nouvelles Microscopies, 45 rue des Saints Pères, Université Paris Descartes, PRES Sorbonne Paris Cité, Paris, F-75006 France.*
[d]*Master Programme: Biologie Cellulaire, Physiologie et Pathologies (BCPP), Université Paris Diderot, PRES Sorbonne Paris Cité, Paris, France.*
[e]*Service Commun de Microscopie (SCM), Institut Fédératif de Recherche en Neurosciences, 45 rue des Saints Pères, Paris, F-75006 France.*

*martin.oheim@parisdescartes.fr*

Evanescent-wave microscopy achieves sub-diffraction axial sectioning by confining fluorescence excitation to a thin layer close to the cell/substrate interface. How thin this light sheet exactly is, however, is often unknown. Particularly in the popular objective-type total internal reflection fluorescence microscopy (TIRFM) configuration large deviations from the expected exponential intensity decay of the evanescent wave have been reported. Propagating, i.e., non-evanescent, excitation light diminishes the optical sectioning effect, reduces contrast and renders the quantification of TIRFM images uncertain. Here, we use a combination of azimuthal- and polar-angle beam scanning, dark-field scatter imaging, and atomic force microscopy to identify the sources of this unwanted background fluorescence excitation. We identify stray light originating from the microscope optics and the objective lens itself as the major sources of background, with minor contributions due to evanescent-wave scattering at the reflecting interface and at refractive-index boundaries in the sample. Apart from evanescence in excitation light, light emitted from a fluorophore can also show observable effects of evanescence. Only fluorophores located close to the coverslip can couple their near-field radiation into propagating waves detectable at supercritical angles. We show that selectively detecting this supercritical-angle fluorescence (SAF) through a high-numerical aperture objective effectively rejects fluorescence from deeper sample regions and improves optical sectioning. The microscopy scheme presented here merges the benefits of TIRF excitation and SAF detection and provides the conditions for quantitative wide-field imaging of fluorophore dynamics at or near the plasma membrane.

It is estimated that 30 to 40% of all cellular proteins reside in the non-aqueous environment of lipid membranes where they perform important metabolic and signaling functions and regulate the transfer of information and material in and out of the cell. Among the techniques used to study membrane dynamics and organization, total internal reflection fluorescence microscopy (TIRFM) occupies a central place. TIRFM is a wide-field technique that confines fluorescence to a thin layer defined by the intensity decay of the evanescent wave set up by total internal reflection of the excitation beam at the cell/substrate boundary. This confinement reduces fluorescence background and photobleaching and is the basis for single-molecule and super-resolution imaging, near-membrane fluorescence correlation spectroscopy, and membrane-selective photoactivation/ photobleaching assays (1-5) all of which rely on an axially well-defined probe volume.

For TIR to occur, light must be directed to the interface at supercritical angles $\theta > \theta_c = \mathrm{asin}(n_1/n_2)$. Here, $n_1$ and $n_2$ are, respectively, the refractive indices of the sample and substrate at the excitation wavelength $\lambda$. In the popular prism-less 'objective-type' configuration (6) a laser beam is focused in the periphery of the back-focal plane (BFP) of a high-numerical aperture objective (NA > $n_2$) and its radial displacement controls the beam angle $\theta$. However, because of local variations of cell adhesion and refractive index ($n_1$) the exact penetration depth $\delta = \lambda/[4\pi(n_2^2\sin^2\theta - n_1^2)^{1/2}]$ of the evanescent wave is often unknown and the interpretation of biological TIRFM images difficult (7-9). Most authors therefore report the calculated values of $\delta$. Additional problems specific to objective-type TIRFM are interference fringes, uneven illumination and contrast degradation due to propagating non-evanescent light that excites fluorescence in deeper sample regions (10-11). The development of techniques that produce better TIRFM images has been the topic of active research in recent years. Azimuthal scanning

TIRFM (12-14) appears well suited to provide more evenly lit TIRFM images, but its actual impact on biological images has not been demonstrated.

Here, we combine rapid acousto-optic polar- ($\theta$) and azimuthal-angle ($\phi$) beam scanning TIRFM (13), dark-field scatter imaging and atomic force microscopy (AFM) to identify sample, coverslip and instrument parameters that contribute to the loss of excitation confinement in objective-type TIRFM and devise strategies for improvement. We show that, if azimuthal beam spinning results in greater image homogeneity, it fails to better confine excitation. The reason is that most diffuse background excitation originates from the beam delivery optics and the objective itself (stray light and high-NA aberrations) rather than from the nanometric roughness of the reflecting interface or scattering at intracellular high-index organelles.

In an attempt to reject this fluorescence background, we combined TIRFM with the selective detection of the directional emission of supercritical angle fluorescence (SAF) (15, 16). Some of the near-field light emitted from near-interface fluorophores converts into light propagating at supercritical angles in a nearby glass substrate. That hollow cone can be captured by the high-NA objectives used in TIRFM, as NA $\geq n_2$. Importantly, none of the far-field radiation is cast into such high angles so that SAF detection suppresses signal from fluorophores located in deeper cytosolic regions. By combining azimuthal beam-scanning EW excitation with SAF detection, we obtain evenly lit images of cultured cortical astrocytes with an improved optical sectioning compared to standard TIRFM images.

# Results

### Beam spinning abolishes image non-homogeneity but does not affect contrast

Optically dense cellular structures, like chromaffin granules (17), or protein-rich adhesion sites spread light by scattering. Evanescent-wave scattering produces a flare of light in the sense of evanescent-wave propagation (17, 18). 'Negative staining' (19) images of an unlabeled BON cell bathed in a fluorescein-containing extracellular solution displayed irregular intensity bands co-linear with the direction of evanescent-wave propagation, **Fig. 1***A*. Changing the azimuthal angle $\phi$ rotated the propagation direction and also changed the stripe orientation. Restoring the illumination symmetry by scanning the spot on a circular orbit with kHz frequency during image acquisition (13) resulted in a more homogenously lit field of view (center image of **Fig. 1***A*). Experimental details are given in **Fig. S1** of the **Supporting Information** (**SI**).

Uneven illumination adversely affects image quality and alters the conclusions drawn from the images. To illustrate this effect, we observed the same cultured cortical astrocytes labeled with FM2-10, a lysosomal marker in these cells (20), with conventional eccentric-spot and spinning TIRFM (spTIRFM), **Fig. 2***A*. While Weber ($C_W$) or Michelson contrast ($C_M$) were unaffected ($C_W$ = 13.6 ± 3.4 for unidirectional vs. 12.3 ± 3.6 for spTIRF; $C_M$ = 0.985 ± 0.006 vs. 0.988 ± 0.008, $n$ = 9 cells, *n.s.*), **Fig.2*B***, more organelles were detected with spTIRFM than with conventional TIRF illumination (66 ± 7 vs. 40 ± 16 spots; 0.018 ± 0.002 $\mu m^{-2}$ vs. 0.010 ± 0.004 $\mu m^{-2}$; $n$ = 9 cells, $p < 0.01$). Individual spots were brighter (4522 ± 1910 cts vs. 2871 ± 2750 cts, mean ± SD, $n$ = 21, $p$ = 0.01) and less variable with spTIRFM than with unidirectional TIRF (42% coefficient of variation, *CV*, vs. 95%), for which lysosomes localized on or alongside the bright excitation bands showed markedly distinct fluorescence, **Fig. 2***C*. With conventional eccentric-spot illumination, such intensity differences would erroneously be interpreted as organelles being located at different axial distances, having unequal dye content or different refractive index. In addition to making intensity measurements more reliable, beam spinning abolished aberrant directional features like a detection bias for mitochondria aligned parallel with the evanescent-wave propagation direction, **Fig. S2**. Denser labeling aggravated the deleterious effects of unidirectional TIRF excitation but independent of the vesicular, mitochondrial, plasma-membrane or cytosolic fluorophore targeting, spTIRFM produced more homogenous images, **Fig. S3**. Restoring the illumination symmetry produces evenly lit TIRFM images, improves the visibility of fluorescent organelles and facilitates the accurate quantification of biological processes at or near the plasma membrane.

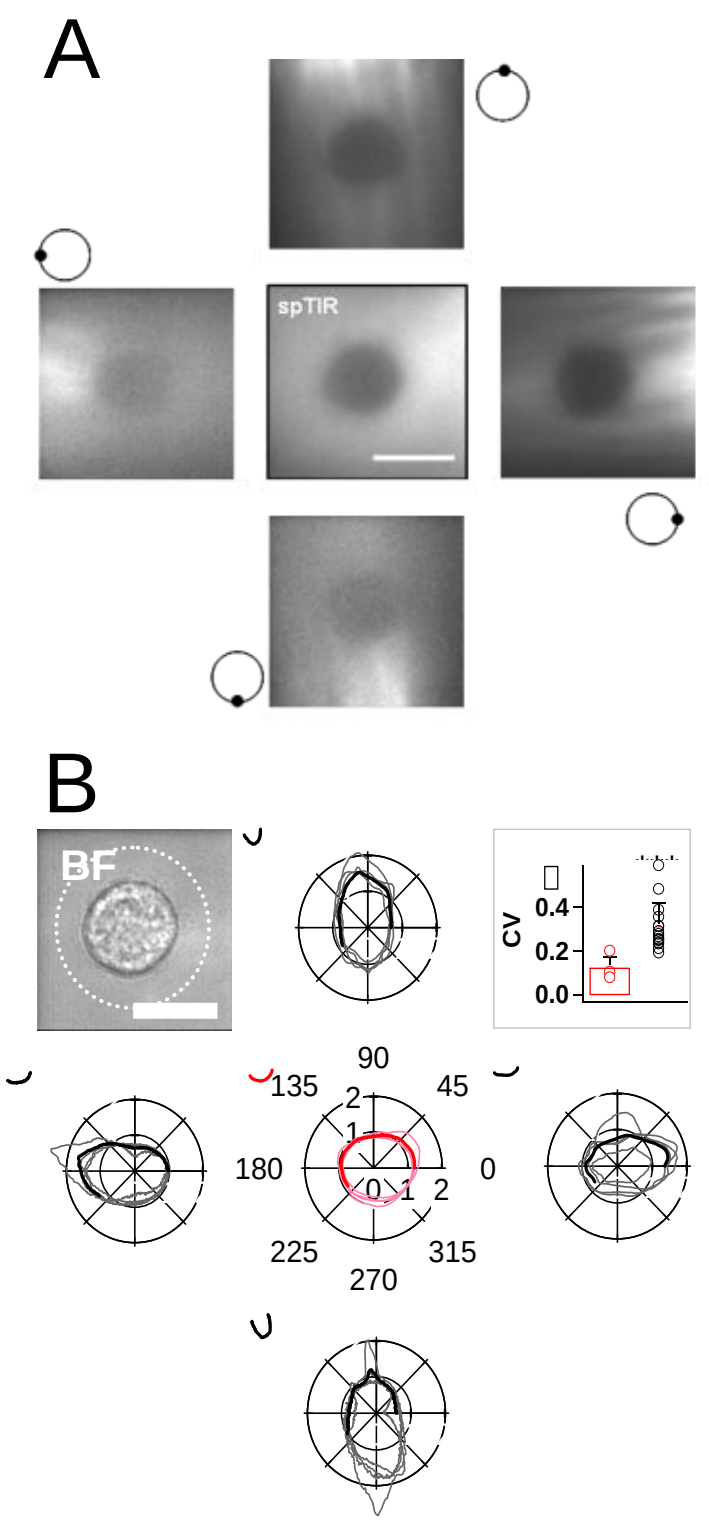

**Fig. 1**. *Azimuthal beam-scanning produces evenly lit TIRFM images. (A), unlabeled BON cell in dye containing extracellular solution. Cardinal images show four standard unidirectional TIRFM images, symbols show position of the focused spot in the objective BFP. Note excitation patterns co-linear with evanescent-wave propagation direction. Rapid azimuthal beam spinning (spTIRFM, center image) evens out the excitation pattern. Polar beam angle $\theta$ was 68°. Scale bar, 10 µm. (B), evolution of intensity along a 2-µm wide circular region (see bright-field image, BF) for different evanescent-wave propagation directions and spTIRFM. Thin lines are individual measurements, solid trace ensemble average over n = 4 cells. Inset shows reduction of the coefficient of variation (CV) of the measured intensity upon spTIRF (0.12 ± 0.05), red, compared to unidirectional TIRFM (black, 0.39 ± 0.15, 0.29 ± 0.05, 0.27 ± 0.03, 0.28 ± 0.14 for NWSE cardinal images, 0.31 ± 0.11 mean ± SD over all directional TIRF images, $p < 0.001$).*

## Beam spinning does not alter image contrast

Compared to the four cardinal images obtained with eccentric-spot illumination, spTIRF images appeared hazier, **Fig. 2***A*. We calculated Weber ($C_W$) and Michelson contrast ($C_M$), two sensitive reporters of image noise and background, respectively. Despite the nominally identical beam angle, $C_W$ and $C_M$ varied among EW propagation directions which was not due to beam misalignment or polarization effects but resulted from the coverslip surface not being perfectly perpendicular to the optical axis. Due to the steep angular dependence of the EW intensity on the polar beam angle $\theta$, even slight (<0.5°) tilt translates to large intensity differences for different azimuthal angles $\phi$. Mounting the sample on a tip-tilt stage allowed us a perfect alignment. Averaged over $n$ = 9 cells, spTIRF did not measurably degrade contrast compared to the average of the four eccentric-spot excitation images but rather eliminated contrast outliers, **Fig. 2***B*, facilitating a cell-to-cell comparison in population studies typical for biological imaging.

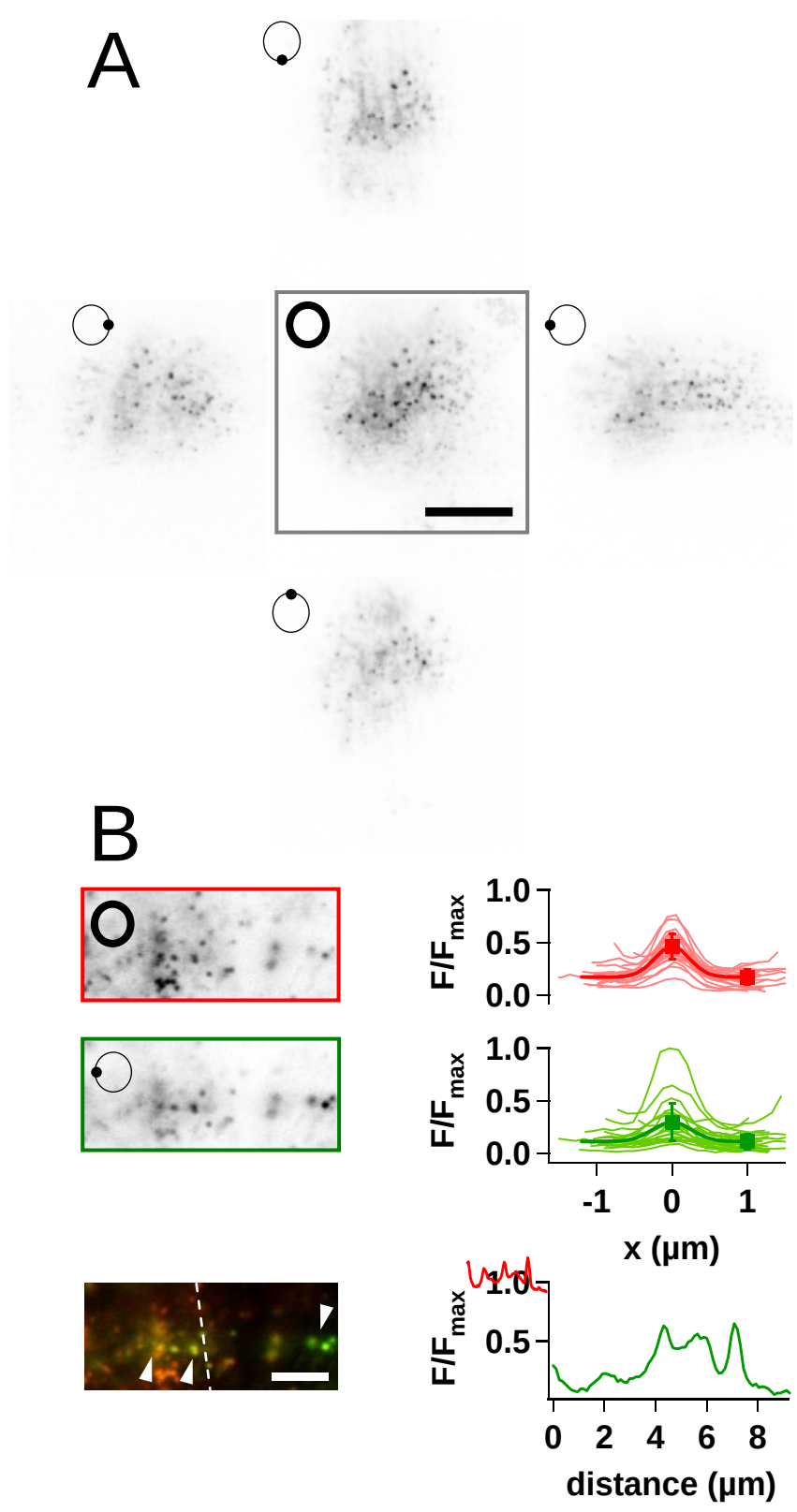

**Fig. 2**. *Uneven illumination affects interpretation of TIRFM images. (A) Unidirectional and spTIRFM images of a FM2-10-labeled mouse cortical astrocyte in culture. Beam angle was 73°. Symbols indicate position of the focused spot in the objective BFP. Note the flare in direction of evanescent-wave propagation absent on the azimuthal beam-spinning image. Contrast inverted for display. Scale bar, 20 µm. (B) Left, detail from a spTIRFM (top) and unidirectional image (middle) and pseudo-color overlay (bottom). Note the intense beam of excitation light propagating across the image (arrowheads). $\theta$ = 70°, scale bar, 5 µm. Right, single-lysosomes intensity profiles upon spTIRFM (red) and unidirectional (green) illumination. Bottom, cross-sectional intensity profile along the dashed line in panel B.*

## Evanescent-wave scattering by the sample plays a minor role

If the observed non-uniformities were primarily sample-induced, shallower penetration depths should result in less scattering (17). Hence, the fluorescence $F_s$' measured in a circular band around an unlabeled BON cell upon spinning excitation (to average out local effects) should decrease with larger beam angles. While $F_s$' decreased, **Fig. 3***A*, the normalized intensity $[F_s'(\theta)]_{norm}$, corrected for the angle-dependence of the evanescent-field intensity $I(\theta)$ itself, was unchanged, **Fig. 3***B*, suggesting that a constant offset rather than scattering at intracellular high-index organelles was the major source of propagated excitation light.

To better separate evanescent and propagating excitation components, we directly quantified scattering by dark-field imaging. A second ×60/NA1.1 water-immersion objective was positioned above the reflecting interface to collect propagating excitation light. No light should enter this objective in the absence of scattering because of total internal reflection. However, even with bare coverslips the mean scattered intensity $I_s$' was non-zero. Dark-field images showed characteristic 'fingerprints' for each objective and beam scanning evened out these patterns without changing the mean intensity, **Fig. 3***C*. Addition of polystyrene latex beads (**Table S3**) changed $I_s$' less than ~10%, although we varied the scattering coefficient $\mu_s$' over five orders of magnitude, **Fig. 3***D*. $I_s$'was slightly larger with bigger beads and at greater penetration depths but overall, we obtained a similar offset for different bead diameters (0.7, 2.8 and 90 µm), at different beam angles ($\theta$ = 64, 70 and 78°), with unidirectional or spinning excitation and for a monolayer of scattering beads on the coverslip surface instead of bead suspensions, **Fig. S4**.

The only slight dependence of non-evanescent excitation light intensities on either the probe volume or scattering coefficient $\mu_s$' confirm that other sources of background dominate over scattering due to sample irregularities.

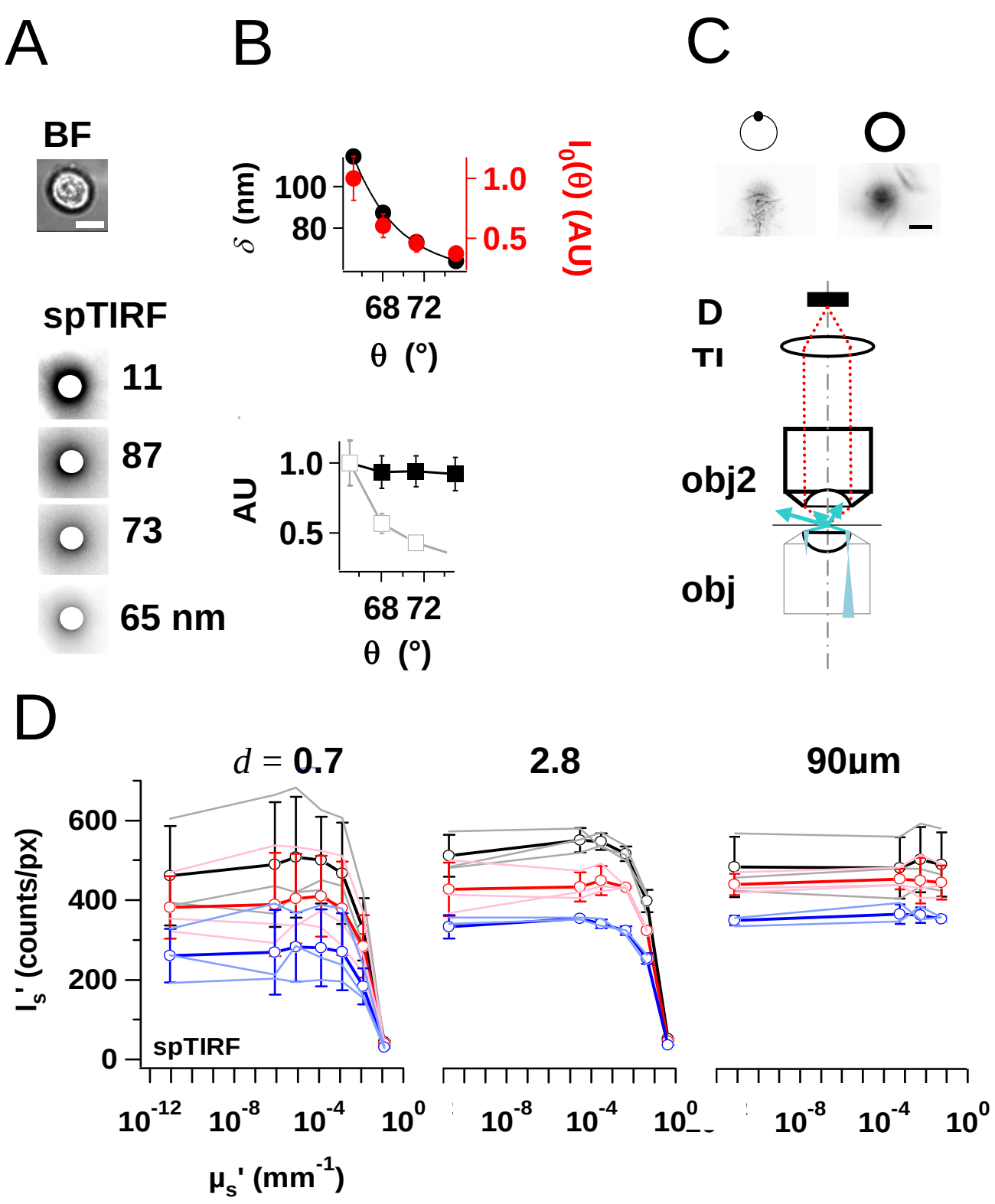

**Fig. 3**. *Sample-induced scattering is not a major source of background. (A), left, bright-field (BF, top) and spinning TIRF (spTIRF, bottom) negative-staining fluorescence images of an unlabeled BON cell for different penetration depths. Images have same grey scale. The footprint region where the cell adheres is blocked out for better clarity. Focus is at the near the glass coverslip. Scale bar, 10 µm. (B), top, dependence on beam angle θ of the calculated evanescent-wave penetration depth δ (black dots) and measured near-surface evanescent-wave intensity I0 (red circles). Bottom, scattered intensity measured in a ring around the cell (open symbols) and normalized with the excitation intensity measured (filled symbols), as a function of θ. Symbols and error bars show mean ± SD from 5 cells. Objective was x60/1.49NA. (C), optical scheme for dark-field imaging. TIRF is set up as before (obj1 – x60/1.45NA or 1.49NA) and scattered light imaged with a second water-immersion objective (obj2 – x60/1.1w, TL – tube lens, D - detector). Dark-field images were obtained a bare BK-7 coverslip upon unidirectional and spinning excitation. (D) Scattered intensity Is' = Is/I0(θ) (corrected for the angle-dependent EW intensity I0(θ)) vs. reduced scattering coefficient µs' for three different beam angles θ (blue - 60°, red - 70° and black - 78°) and three different diameters of scattering beads (from left to right: 0.7, 2.8 and 90 µm) upon 488-nm spinning excitation. Light lines are individual measurements, dark lines, symbols and error bars show means ± SD from triplicate experiments. Inner filtering due to multiple scattering results in a drop of Is' at high µs'.*

## Coverslips are rough on the length-scale probed by TIRF

TIRF produces an excitation maximum at the reflecting interface and therefore is very sensitive to surface irregularities. Perhaps these are more important than volume scattering. Direct evidence for the relevance of surface scattering comes from the observation of non-specular reflections of slow atoms in evanescent-wave mirrors that were abolished by flame-polishing the substrate (21).

AFM images revealed scratches, dimples, and holes on some borosilicate coverslips but no obvious defects on others; quartz had a 'rolling hill' aspect, **Fig. 4***A*. Rejecting coverslips with large irregularities, the remainder had sub-nanometric RMS roughness ($R_q$), including after coating with polyornithine or collagen, **Fig. 4***B*. However, $R_q$ increased to ~2 nm when wetting coverslips, with peak roughness $R_p$ of tens of nanometers, and peak-to-peak heights up to 60 nm, **Table 1**. Thus, under biological recording conditions, cell adhesion molecules produce height features comparable with the penetration depth of the evanescent wave.

**Table 1:** *Coverslip surface roughness*

| | *BK-7* | | *quartz* | |
|---|---|---|---|---|
| | bare, dry (*n* = 8) | | bare, dry (*n* = 4) | |
| $R_q$ (nm) [a] | 0.24 ± 0.04 | | 0.37 ± 0.05 | |
| $R_q$ (nm) [a] | polyornithine, dry (*n* = 5) | collagen, dry (*n* = 5) | polyornithine, dry (*n* = 6) | collagen, dry (*n* = 4) |
| | 0.44 ± 0.12 | 0.44 ± 0.27 | 0.65 ± 0.31 | 0.69 ± 0.06 |
| | polyornithine, wet (*n* = 10) | | polyornithine, wet (*n* = 5) | collagen, wet (*n* = 5) |
| $R_q$ (nm) [a] | 5.4 ± 2.2 | | 2.8 ± 1.2 | 2.6 ± 0.3 |
| $R_q$ (nm) [b] | 1.4 ± 0.5 | | 2.8 ± 1.2 | 2.0 ± 0.4 |
| $R_a$ (nm) [b] | 1.0 ± 0.3 | | 1.6 ± 0.8 | 1.1 ± 0.2 |
| $R_t$ (nm) [b] | 24.1 ± 8.4 | | 27.8 ± 14.6 | 49.8 ± 9.1 |
| pk-pk height (nm) [b] | 16 (min); 60 (max) | | 13 (min); 48 (max) | 41 (min); 62 (max) |

[a] RMS roughness over 5μm × 5μm of non-selected commercial coverslips.
[b] RMS and absolute roughness as well as peak height, calculated over 2μm × 2μm regions of interest from selected coverslips devoid of large surface defects.

## Glare and aberrations of peripheral beams limit excitation confinement

Variable-angle or beam-spinning TIRFM uses 'critical illumination' in which the surface of a rotating wedge (12), a scan mirror (14) or a pair of AODs (2, 13) is imaged into the sample plane, **Fig. 5***A*. As a consequence, any irregularities, dust or scratches on the scanning device or a conjugate plane are imaged into the sample plane and produce stray light. Microscope-induced glare was measured in dark-field by moving the upper objective: a defocus by *dz* displaces the conjugate image plane located inside the microsope by *dz* multiplied by the longitudinal magnification (i.e., the lateral magnification squared). On the plot of $I_s$' vs. *dz* the coverslip and AOD surfaces are easily recognized as peaks of the dark-field signal, **Fig. 5***B* (black trace). Placing an appropriately-sized disk in a conjugate aperture plane of the excitation path blocked most of this stray excitation (red trace).

A second source of non-evanescent excitation light are stray reflections from inside the objective. TIRFM uses its extreme periphery for guiding the excitation beam at supercritical angles to the reflecting interface. To assess beam quality beyond beam angles otherwise obscured by TIR, we imaged the beam with an oil-coupled high-index solid immersion lens onto a wave front analyzer and calculated the beam parameter product ($M^2$) and Strehl ratio as a function of $\theta$, **Fig. 5***C* Two objectives from different manufacturers showed stable beam quality up to $\theta$ ~ 50° that then deteriorated abruptly, indicating that off-axis phase aberrations or partial obstruction degrade objective performance. These effects could be due to a NA smaller than specified. Early TIRF microscopists remember that some lenses sold as 1.4-NA actually had an effective NA ($NA_{eff}$) closer to 1.38. Measuring $NA_{eff}$ using a technique based on supercritical angle fluorescence (SAF) detection (22), **Fig. S5**, we found that the1.45–1.46 objectives generally fulfilled their specification, but a nominal 1.49-NA lens only had $NA_{eff}$ = 1.47, **table 2**.

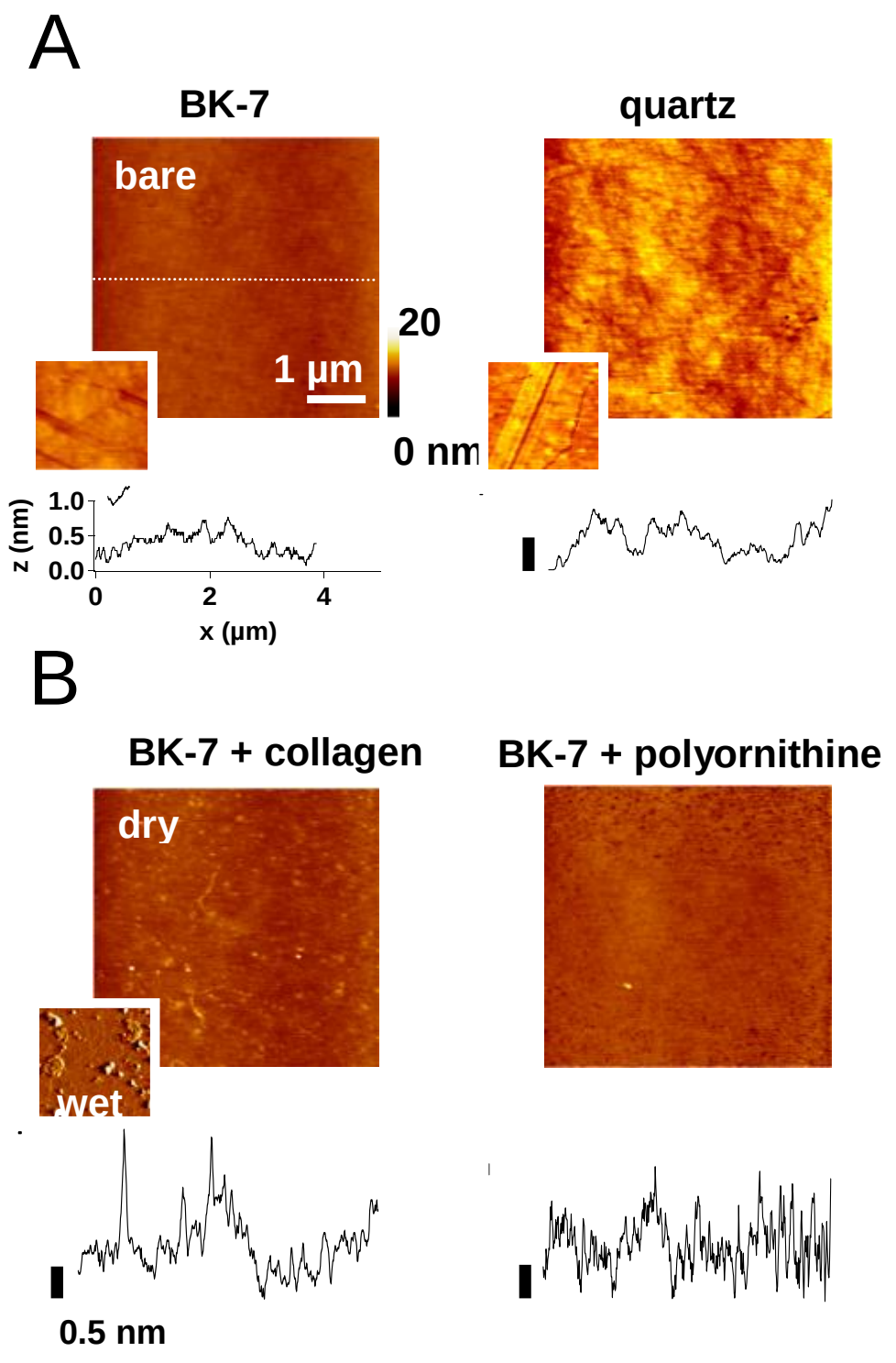

**Fig. 4**. *Coverslips display height variations relevant on the length-scale probed by TIRFM. (A), examples of contact-mode atomic-force microscopy (AFM) images showing the surface roughness of bare BK-7 (left) and quartz coverslips (right). Vertical tip displacement (i.e., surface height) is pseudo-color coded from 0 to 20 nm. The RMS roughness Rq was 0.43 and 0.58 nm, respectively, for the images shown. Insets show scratches and surface defects of coverslips that were rejected. Curves show surface profiles along the dotted line as shown on the left image. (B) Same for BK-7 after collagen (left) and polyornithine treatment (right), seen as a fibrous deposit. Rq were 0.49 and 0.38 for the images shown respectively. Inset, tapping-mode AFM image illustrating the marked morphology change of collagen upon hydration. Note that height scale is 0 to 200 nm for inset. See table 1 for quantifications.*

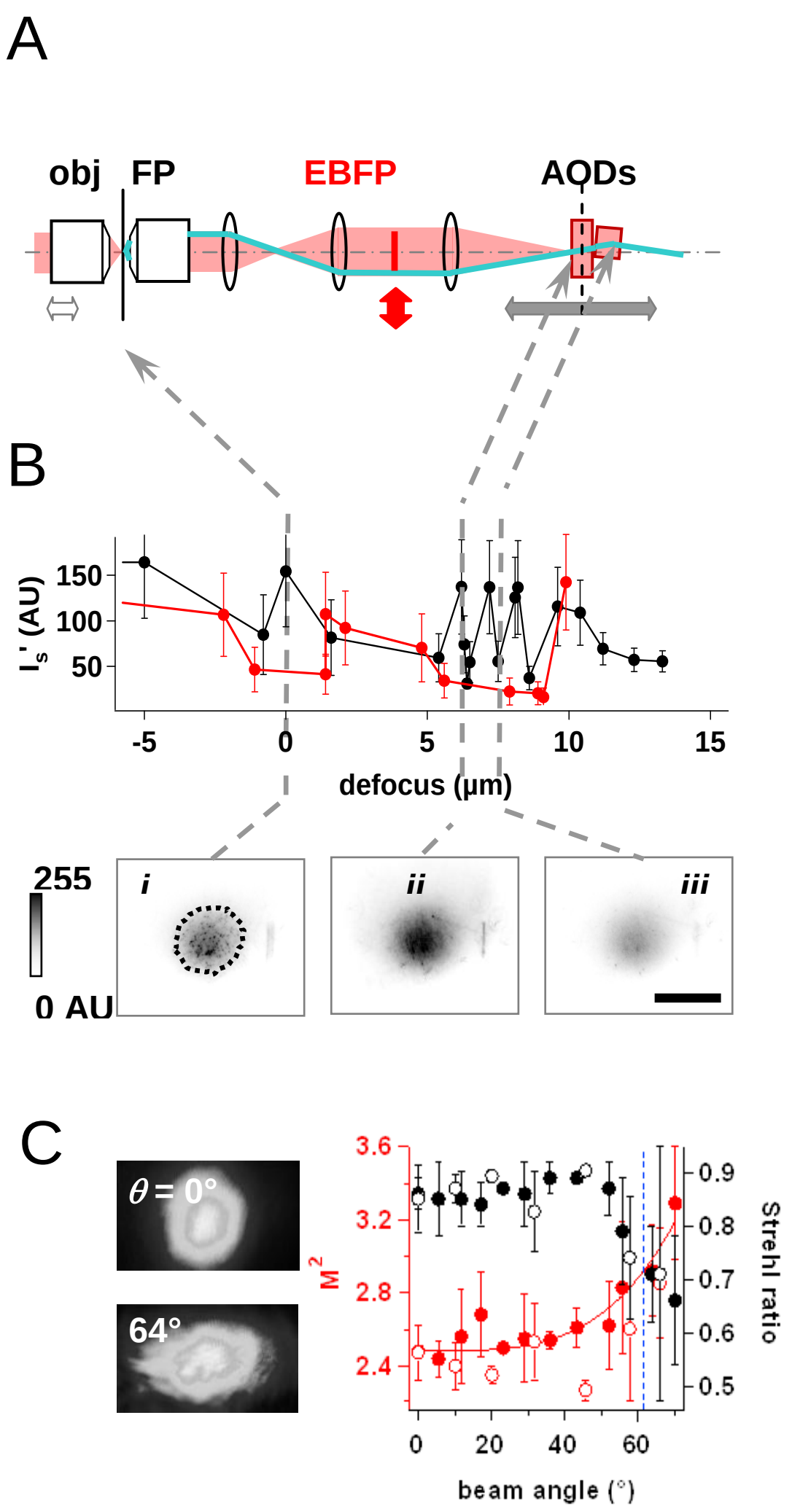

**Fig. 5**. *Glare and off-axis aberrations limit excitation confinement. (A), dark-field detection of microscope-induced stray light Is'. Scanning (white arrow) the focal plane (FP, solid line) of the upper objective (obj) moves its conjugate focal plane (dashed) across the excitation optical path (grey filled arrow), allowing to identify the sources of stray light. (B), top, Peaks of Is' were detected (black trace) at the coverslip surface (i), at consecutive AOD surfaces (see ii as an example), and - to a lesser degree - within the AOD crystal, (iii), and are reduced when placing an opaque disk [red on (A)] in the center of an aperture plane (equivalent back-focal plane of the objective, EBFP), red trace. Bottom, representative dark-field images taken at planes indicated. Dashed line identifies region of intensity measurements. Scale bar, 25 µm. (C), left, example cross-sectional intensity images (top) taken at a beam angle $\theta$ of 0° and 64°, with $M^2$ and Strehl ratio values of 2.5 and 0.8 (left) and 3 and 0.6 (right), respectively. Bottom, evolution of $M^2$ (red) and Strehl ratio (black) with $\theta$, for the x60/NA1.45 (solid symbols) and x100/NA1.46-objective (circles). Symbols and error bars represent mean ± SD from triplicate measurements. Note the degradation of beam quality starting well before the critical angle $\theta c$ for the BK-7/water interface (dashed, blue).*

**Table 2:** *Effective NAs of TIRFM objectives*

| | $NA_{eff}$ |
|---|---|
| PlanApo ×60/NA1.45 (objective 1) | 1.468 ± 0.004 [a] |
| PlanApo ×60/NA1.45 (objective2) | 1.456 ± 0.003 |
| αPlan-Apochromat ×100/NA1.46(objective 1) | 1.453 ± 0.004 |
| αPlan-Apochromat ×100/NA1.46(objective 2) | 1.457 ± 0.007 |
| ApoN ×60/NA1.49 | 1.474 ± 0.007 |

[a] Measurements (ref. (22)) were robust against re-aligning the Bertrand lens, re-focusing or moving laterally in the fluorophore layer. Absolute variations are given as uncertainties.

Taken together, our experiments identify the objective as a non-negligible contributor to non-evanescent excitation light in prismless TIRFM. While permitting near-membrane imaging on a slightly modified epifluorescence microscope, through-the-objective TIRFM inevitably comes at the price of a degraded excitation confinement. Objective-induced glare is a consequence of the extreme off-axis use of the objective and, as such, it is the same for unidirectional vs. spTIRF. Thus, other strategies than beam spinning must be devised to abolish the long-range excitation component introduced by the microscope objective.

### Supercritical angle fluorescence (SAF) detection rejects nonevanescent background

Fluorophores closer than a light wavelength from the coverslip change their radiation pattern because evanescent emission components couple to the surface and become propagative. These waves are directed exclusively into NAs > $n_2$, i.e., angles beyond the critical angle, which is the same as it would be for TIR excitation at the same wavelength. None of the fluorescence originating from deeper within the sample can be emitted in this hollow cone. SAF microscopy discriminates between surface-proximal and distant fluorophores by selectively collecting this high-NA information and leads to a similar axial confinement as EW excitation (23, 24). Undercritical angle fluorescence (UAF) is typically rejected by obstruction of the central part of the objective pupil. This, together with high-NA polarization effects degrades resolution (15, 16). An elegant solution is the subtraction of an image acquired at a slightly lower NA (collecting only UAF, **Fig. 6***A*), from the image taken at full aperture (collecting UAF + SAF) that creates a 'virtual' SAF image $I_{vSAF} = I_{UAF+SAF} - I_{UAF}$ (25), **Fig. 6***B*. Because high NAs are used for the acquisition of both images, resolution is preserved, **Fig.6***C* and **Table S1**. We then compared spTIRFM images of cortical astrocytes transfected with vinculin-GFP, a membrane-cytoskeletal protein involved in the linkage of integrin to actin. We adjusted the beam angle to 74°, corresponding to a calculated $\delta$ of 72 nm. Both UAF and UAF + SAF images taken with a ×100/1.46-NA objective showed the characteristic focal adhesion sites on the substrate (**Fig. 6***C*) but the *v*SAF image was devoid about half of the vinculin-bearing vesicles and tubules seen on the conventional TIRFM image, indicating that these were located close enough to the reflecting interface to be reached by excitation light, but too far from the reflecting interface to emit SAF. Compared classical full-NA collection, SAF detection captures about one third of the fluorescence. Nevertheless, the suppression of background fluorescence outweighs the signal loss, because the SAF image contains information otherwise obscured by background. An example is shown in **Fig. 6***D*, where SAF reveals an alternate sequence of membrane-proximal and distant stretches rather than a single flat adhesion site as seen in TIRFM, **Fig. 6***D*. We conclude that the combined use of evanescence in excitation and emission presents clear advantages for imaging fluorophores near or at the basal plasma membrane compared to TIRFM alone. *v*SAF detection has the extra benefit of abolishing the background resulting from non-evanescent excitation that is notorious with objective-type TIRFM.

## Discussion

The use of supercritical-angle fluorescence excitation and emission collection both date back to the 1960-70's. Their combination in confocal-spot TIRFM with a parabolic mirror (26), or the combination of prism-based excitation and SAF-detection through a high-NA objective (27) produces important sensitivity gains in single-molecule detection. We now show that simultaneous supercritical excitation and emission collection overcomes the major inconvenience of objective-type TIRFM and allows more reliable fluorophore localization.

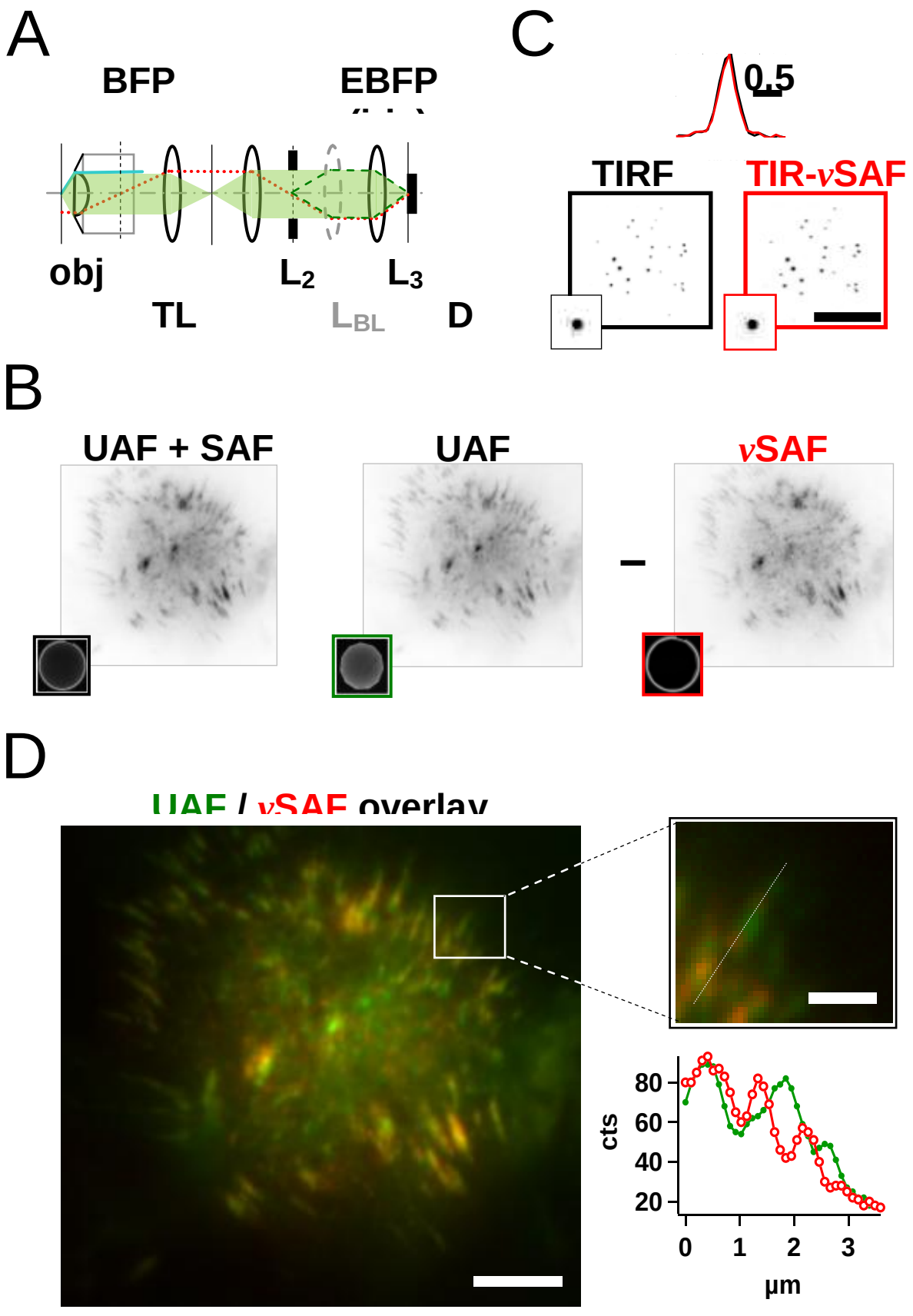

**Fig. 6**. *Virtual super-critical angle fluorescence (vSAF) detection improves axial confinement in objective-type TIRFM. (A) Schematic layout of the emission optical path. Sample fluorescence (green) is imaged via the objective (obj) and tube lens (TL) onto the detector (D). Lenses L2 and L3 create an intermediate aperture plane (equivalent back focal plane, EBFP) permitting the selection of emission angles (Fourier-plane filtering) by an adjustable iris. A removable Bertrand lens (LBL) permits imaging the EBFP for alignment. (B), spTIRFM images of 93-nm red-fluorescent microspheres (inset shows single bead) and intensity profile upon conventional full-aperture detection (TIRF) and after vSAF detection, i.e., subtraction of an image taken with the iris adjusted to collect only undercritical angle fluorescence (UAF). Lateral resolution is identical (see Table S1) because high NAs are used for the acquisition of both images. (C), field- and aperture-plane (insets) images of a cultured cortical astrocyte expressing vinculin-EGFP. Autoscaled images show, from left to right, spTIRFM image acquired at full objective NA (i.e., collecting both UAF and SAF), the UAF image with the iris partially stopped down, and the resulting 'virtual' SAF (difference) image, containing only SAF. (D), green/red pseudocolor overlay of the UAF and vSAF images as in (C). Inset shows boxed region and line profile along dotted line, showing fine detail on the vSAF image.*

Conventional eccentric-spot TIRF produces non-homogeneities and directionality that can be avoided by beam spinning (**Figs 1**, **2**, **S2** and **S3**), but this does not fundamentally address the problem of spurious far-field excitation (11). The reason is that most of the diffuse background in prismless TIRFM stems from the objective and beam-delivery optics and is independent of the evanescent-wave penetration depth (**Fig. 3**) or propagation direction (**Fig. S4, 5**). A minor contribution is due to scattering at refractive-index heterogeneities in the sample (**Fig.3**) and irregularities of the reflecting interface (**Fig.4**). By selecting for surface-proximal fluorophores SAF detection suppresses this background, improving optical sectioning and image contrast. Concurrent spTIR-*v*SAF imaging outperforms conventional TIRF for near-membrane fluorophore detection (**Fig.6**).

Compared to the alternative trans-prism geometry, objective-type TIRFM (6) produces brighter but less crisp images (10, 23, 28). Brightness is a result of high-NA SAF collection (23, 29). Reduced contrast has been attributed to stray reflections (4, 11, 13), interference (10, 30-32), or propagating shafts of light emerging from the margin of the objective (31). Moerner and colleagues estimated that as little as 26% of the initial power is returned from the objective in the 'totally' reflected beam (33), suggesting important losses along the excitation path (34). Identifying their origin has been difficult because of the restricted access inside the microscope. We now identify glare and partial obstruction of the beam at high NAs as the major sources of non-evanescent excitation.

Two strategies have been used to improve excitation confinement. Two-photon fluorescence excitation in TIRFM reduces background because scattered photons are too dilute to result in appreciable fluorescence generation. Although effective, the technique has been little used (20, 36). Alternatively, combining one-photon excitation with a continuous change of the direction of EW propagation (11-13, 32) results in a time-averaging over non-homogeneities. Both approaches are analogous to those used in selective-plane illumination microscopy (SPIM) (37, 38).

We show that spTIRFM is superior to classical TIRFM for quantitative near-membrane imaging, but it is no cure-all against non-evanescent excitation light. Because the scanning device is imaged into the sample plane, any imperfection appears in the field-of-view. Fourier-plane filtering reduces background, but does not affect stray light generated in the excitation path downstream of the aperture mask, arising from the passage through the periphery and clipping of the excitation beam inside the microscope objective. We show that this stray light is efficiently rejected by SAF detection. Again, an analogy can be drawn with SPIM for which a combination of scanning light-sheet illumination and confocal slit-scanning detection produces crisper images (38).

Our study underpins the importance of testing high-NA objectives individually prior to order. Not all objectives met their specifications suggesting that the commercial race to higher NAs has led to somewhat overly optimistic statements of performance. Also, all objectives displayed measurable autofluorescence that interfered with the detection of faint green and yellow fluorescence (**Fig. S5**), which is the reason why we used red-emitting beads for PSF measurements.

How does TIR-*v*SAF compare to alternative approaches that increase the signal-to-background ratio? With a NA-1.46 objective, *v*SAF collects about one third of the emission of a fluorophore located within 100 nm from the interface, compared to conventional full-aperture detection. The calculation of the *v*SAF image requires $N$+1 planes and thus produces lower phototoxicity/bleaching compared to TIRF deconvolution, as soon as three or more z-planes are acquired. Similar to surface-plasmon enhanced fluorescence (39, 40), spTIR-*v*SAF achieves optical sectioning through its distance-dependent collection efficiency, however, neither does it rely on special silver-coated coverslips, nor does it suffer from metal-induced quenching of near-surface fluorophores. *v*SAF detection requires only a minor modification of the collection optical path. It can be implemented on any standard microscope where it will improve single-molecule detection, TIR-FCS or TIRF-FRAP measurements, and the growing number of super-resolution techniques including standing-evanescent-wave structured illumination, TIRF-STED and localization-based (PALM/STORM) microscopies. The association of TIRFM and *v*SAF is of particular interest with the recent ×100 and ×150 objectives that, as a consequence of their small back pupils, impose a very precise control and tight focusing of the excitation beam in the BFP.

## Materials and Methods

**Cell culture.** Astrocytes and BON cells were cultured as described in the SI. FM2-10 and FM4-64 labeled astrocyte lysosomes (18). Cells were transfected with plasmids encoding VAMP2-EGFP, mito-EGFP, Lck-EGFP, vinculin-GFP or CD63-GFP. Experiments were performed at 22-23°C.

**spTIRF microscopy.** A 488-nm laser spot was scanned in the objective BFP (13) under LABVIEW control (National Instruments) to adjust the polar (□) and azimuthal angle ( ) of the totally reflected beam, (Fig. S1). Objectives (Table 2) were piezo-mounted for accurate focusing (PIFOC, Physik Instrumente). Fluorescence was detected through filters listed in Table S4 on electron-multiplying charge-coupled device cameras (EMCCD, Photometrics). Total magnifications were 80, 103 and 120 nm/pixel, as indicated. We used low µW powers in the sample plane.

**SAF detection and effective-NA measurements.** For SAF, an iris was positioned in an EBFP of the objective while the sample was imaged on the EMCCD. An equally removable Bertrand lens permitted EBFP imaging. Both the aperture mask and lens could be centered with precision translation tables. $NA_{eff}$ was measured using a FITC (500 µM) spin-coated coverslip (**Fig. S6**). A MATLAB (Mathworks) routine allowed determining $r_c$ and $r_{NA}$ from the circularly averaged steepest intensity change. $r_{NA}$ and $r_c = f \cdot \sin(\square_c)$ are, respectively, the measured intensity cut-offs due to the upper limit of the objective effective NA collection angle and the radius for which TIR at the air/substrate interface occurs. $NA_{eff} = r_{NA}/r_c$.

**Darkfield scatter imaging.** Scattering coefficients of bead suspensions (Polysciences, **Table S3**) were calculated using the interactive Mie Scattering Calculator, http://omlc.ogi.edu/calc/mie_calc.html written by Scott Prahl. Scattered excitation light was detected through a dipping objective (LUMFl ×60/NA1.1, Olympus).

**AFM.** Images were obtained on a Bioscope (Veeco) with Nanoscope IIIa controller. Surface roughness was evaluated by the calculation of $R_q = \sqrt{\frac{1}{MN}\sum_{j=1}^{M}\sum_{i=1}^{N} z^2(x_i, y_j)}$, $R_a = \frac{1}{MN}\sum_{j=1}^{M}\sum_{i=1}^{N} \left| z(x_i, y_j) \right|$ and $R_t = \max\left(\left|z_{max} - z_{min}\right|\right)_{\forall(x,y)}$. $M$ is the number of

points per scan line and $N$ is the number of lines, $z(x,y)$ is the vertical tip displacement at point $(x,y)$. Tapping-mode AFM was used for wet coverslips.

**Image analysis.** Dark images were subtracted from all fluorescence images. Weber contrast was calculated as $C_W=(I-I_b)/I_b$ ,where $I$ and $I_b$ are the fluorescence intensity of image features and background, respectively. $I_b$ was measured in a large cell-free polygonal ROI identified on the bright-field/fluorescence images. Signal intensities $I$ were measured in ROIs outlined by a border exceeding two times the SD of $I_b$. Michelson contrast (visibility) was calculated as $C_M=(I_{max}-I_{min})/(I_{max}+I_{min})$ , RMS contrast as $C_{RMS}=\sqrt{\frac{1}{MN}\sum_{j=0}^{M}\sum_{i=0}^{N}(I_{ij}-\bar{I})^2}$ , where $I_{ij}$ is the intensity of pixel $(i, j)$ of a $M \times N$ pixel image. $\bar{I}$ is the average fluorescence of all pixels. Line-profiles across single lysosomes measured on spTIRF and unidirectional TIRF images were compared by fitting the central intensity peak with a Gaussian distribution and noting peak and local background fluorescence. Image analysis was performed with MetaMorph (Molecular Devices) and IGOR (Wavemetrics) using custom macros.

**Acknowledgment.** We thank K Hérault and C Debaecker for cell culture, D Li, M van 't Hoff and C Ventalon for help with experiments/programming, P Jegouzo for custom mechanics, F Jean for administrative support and V Emiliani, O Faklaris (IJM Paris), S Konzack (Olympus), G Louis, H Schrader (Zeiss) for the loan of equipment. We thank T Barroca, F Lison, C Pouzat and R Uhl for discussion, RH Chow, D Li, N Ropert and C Ventalon for comments on earlier versions of the manuscript and JS Kehoe for careful proofreading. Financed by the EU (FP6-STRP-n°037897-AUTOSCREEN, FP7-ERA-NET n°006-03-NANOSYN), the *Agence Nationale de la Recherche* (ANR P3N 09-044-02 nanoFRET²), the FranceBioImaging (FBI) initiative and mobility support from the Franco-Bavarian University Cooperation Centre (BFHZ-CCUFB). AFM imaging was performed on the Paris Descartes St Pères core imaging facility, *Service Commun de Microscopie* (SCM).

# SUPPORTING INFORMATION

## List of abbreviations

| | | |
|---|---|---|
| AFM | - | atomic force microscopy |
| AOD | - | acousto-optical deflector |
| AOTF | - | acousto-optical tunable filter |
| aq. | - | aqueous |
| BF | - | bright field |
| BK | - | borosilicate (glass) |
| BON | - | A cell line derived from a metastatic pancreatic human carcinoid tumor |
| CNRS | - | Centre National de la Recherche Scientifique |
| DAC | - | digital-to-analog card |
| (E)BFP | - | (equivalent) back focal plane |
| (EM)CCD | - | (electron-multiplying) charge-coupled device |
| EtOH | - | ethyl alcohol, ethanol |
| EU | - | European Union |
| EW | - | evanescent wave |
| FITC | - | fluorescein isothiocyanate |
| FOV | - | field of view |
| IMA | - | Integrated Morphometry Analysis |
| IQR | - | interquartile range |
| NA | - | numerical aperture |
| *n.s.* | - | not significant |
| OPSL | - | optically pumped solid-state laser |
| PALM | - | photoactivated localization microscopy |
| pk-pk | - | peak-to-peak |
| PSF | - | point spread function |
| RMS | - | root mean square |
| ROI | - | region of interest |
| RT | - | room temperature, 22-23°C |
| SAF | - | supercritical angle fluorescence |
| SIL | - | solid immersion lens |
| spTIRF | - | spinning TIRF |
| STED | - | stimulated emission depletion microscopy |
| STORM | - | stochastic optical reconstruction microscopy |
| TIR[F(M)] | - | total internal reflection [fluorescence (microscopy)] |
| UAF | - | undercritical angle fluorescence |
| VCO | - | voltage-controlled oscillator |
| *v*SAF | - | virtual supercritical angle fluorescence |

# SI Methods

**Coverslips, dyes and bead suspensions.** 25-mm diameter #1 and #1.5 coverslips (147.8 ± 2.8 µm and 173.6 ± 2.8 µm thick, respectively, *n* = 35 each, SchottDesag D263M, Thermo Fisher Menzel, Braunschweig, Germany) were sequentially passed twice through baths of 70% EtOH and sterile water, respectively, and were used thereafter or else treated with poly-ornithine (1.5 µg/ml, 30 min, 37°C, 5% $CO_2$) or collagen (Glassand rat tail acid soluble Bornstein Traub type I collagen, Sigma, Lyon, France). Collagen was prepared at 1 mg/ml in aq. 1% acetic acid. This stock was diluted 1:200 in 30% EtOH and the coverslips incubated 3h at room temperature (RT, 22-23°C), the excess liquid removed and the coverslips dried (30 min, RT) under dust-free air flow before use. Coverslips were stored individually in sealed 6-well plates. Surface roughness of bare and treated BK-7 and fused silica substrates (TGP Inc., Painesville, OH) was measured with AFM (Bioscope, Veeco, Plainview, NY).

Submicron layers of rhodamine 6G (R6G, Sigma) or fluorescein isothiocyanate (FTIC, Fluka, Buchs, Switzerland) were deposited on coverslips with a spin coater (KW-4A, SPI supplies, West Chester, PA) at 3,000 rpm. Dilute solutions of red-emitting polystyrene latex microspheres (488/685 nm TransFluoSpheres, 93-nm diameter, Invitrogen, Saint Aubin, France) were drop-cast onto a clean coverslip and immobilized by solvent evaporation and used for the measurement of point-spread functions (PSFs). The pixel size for PSF measurements was 77 nm. Refractive indices $n_D^{22-23}$ of immersion liquids were measured (589 nm, RT) with an Abbe refractometer (WYA, Shanghai, China; **Table S2**). Suspensions of non-fluorescent PS latex beads (Polysciences, Inc., Warrington, PA, **Table S3**) were used as scattering samples. Scattering properties were calculated using the Mie Scattering Calculator at http://omlc.ogi.edu/calc/mie_calc.html written by Scott Prahl, assuming $n_{PS}$ = 1.6053 at 488 nm, from http://refractiveindex.info by Mikhail Polyanskiy.

**Cell preparation.** Experiments followed EU and institutional guidelines for the care and use of laboratory animals (Council directive 86/609EEC). Astrocytes were prepared from P0-1 (P0 being the day of birth) NMRI mice (Janvier, Montpellier, France) as described (3). Briefly, neocortices were dissected and mechanically dissociated. Cells were plated and maintained in Petri dishes for one week to reach confluence before their transfer onto poly-L-ornithine-coated cover slips. Secondary cultures were maintained in DMEM, supplemented with 5% FCS, penicillin (5 U/ml), and streptomycin (5 µg/ml) at 37 °C in a humidified 5% $CO_2$ atmosphere. Astrocytes kept for one more week in secondary culture and incubated with the fluorescent lysosomal marker FM2-10 (50 µM, 30 min) (4) or transfected with the plasmids listed in **Table S4** were used for imaging, during which they were continuously perfused at 1 ml/min with extracellular saline containing, in mM: 140 NaCl, 5.5 KCl, 1.8 $CaCl_2$, 1 $MgCl_2$, 20 glucose, 10 HEPES (pH 7.3, adjusted with NaOH). Isolated astrocytes or small islets of astrocytes were imaged after 20 min wash (FM dyes) or 24-36 h following transfection.
BON cells were a gift from Dr C. Desnos (CNRS UMR8192, University Paris Descartes) and were cultured and plated as described (5). The BON cell line was established from a lymph node metastasis of a human pancreatic carcinoid tumor (6) and provided to CNRS by Dr C. M. Townsend (University of Texas, Medical Branch, Galveston, TX). Cells were cultured at 37°C in a humidified 5% $CO_2$ atmosphere in Ham's F-12/DMEM with 10% fetal bovine serum and seeded onto collagen-coated glass coverslips for imaging, 1-5 days after plating.

**Azimuthal beam-spinning TIRFM**. In objective-type TIRFM, a laser spot is focused in an eccentric position in the BFP of an objective with $NA = n_2 \cdot \sin(\theta_{NA})$, producing an oblique collimated beam emerging from the objective at an angle

$$\theta = \arcsin\left[M \cdot r / (n_2 \cdot f_{FL})\right]. \quad \text{(eq.1)}$$

The reflection is total for radii $r \geq r_c = n_2 \cdot f$, implying that $NA \geq n_1$. Here, $r$, $f_{FL}$ and $M$ are the spot's radius from the optical axis, the focal length of the focusing lens and the objective transverse magnification, respectively. For an aplanatic objective, the lateral magnification is $M = f_{TL} / f_{obj}$, where $f_{TL}$ is the focal length of the used tube lens.
We modified an earlier described microscope (7), **Fig.S1***A*. Briefly, the 488-nm line of an $Ar^+$-ion laser (Reliant 150, LaserPhysics, West Jordan, UT) or 488-nm OPSL (Sapphire 488, 20 mW, Coherent, Santa Clara, CA) was isolated and shuttered with an AOTF (AA.Optoélectronique, St.Rèmy-en-Chèvreuse, France), spatially filtered by passing it through a mono-mode optical fiber (kineFLEX, Qioptiq, formerly Point Source, Hamble, Southhampton, UK) and expanded by an achromatic ×2.8 telescope ($f_1$ = 50 mm, Thorlabs, Dachau, Germany; $f_2$= 140 mm, Qioptiq, formerly Linos, Göttingen, Germany). Polarization was linear.

Two crossed AODs (AA.DTS XY-250@405nm, AA.Optoélectronique) were positioned ($x,y,z,\theta,\phi$) in a conjugate field plane. A compressing telescope (×0.25, $f_1$ =200 mm, $f_2$ = 50 mm, Linos) increased the scan angle. An appropriately sized iris and opaque disc were placed in the conjugate aperture plane (EBFP of the objective) to reduce stray light from the AOD surfaces. A high-quality focusing lens (Rodagon $f$ = 135 mm, Qiopiq, formerly Rodenstock, Feldkirchen, Germany) and high-NA oil-immersion objective ($f$=3 mm, PlanApo ×60/NA1.45oil TIRFM, or APO N ×60/NA1.49oil (Olympus, Hamburg, Germany) formed another telescope (×0.022), resulting in a circular on-axis Gaussian beam of ~ 21 ± 2 µm (30 ± 1 µm) FWHM ($1/e^2$) diameter in the sample plane and 0.77° ± 0.07° (13 ± 1 mrad) far-field divergence (half angle of the FWHM intensity) emerging from the objective. This optical path allowed us to freely position within µs the focused spot in the BFP up to 5 mm off the optical axis, a radius larger than the pupil diameter of the NA1.49 objective $r_{NA}$= NA·$f$ = 4.47 mm). For SAF experiments, we used a αPlan-Apochromat ×100/NA1.46oil (Zeiss, Oberkochen, Germany). The objective position could be adjusted with a piezo-electric focus drive (P721.PLQ, Physik Instrumente, Karlsruhe, Germany).

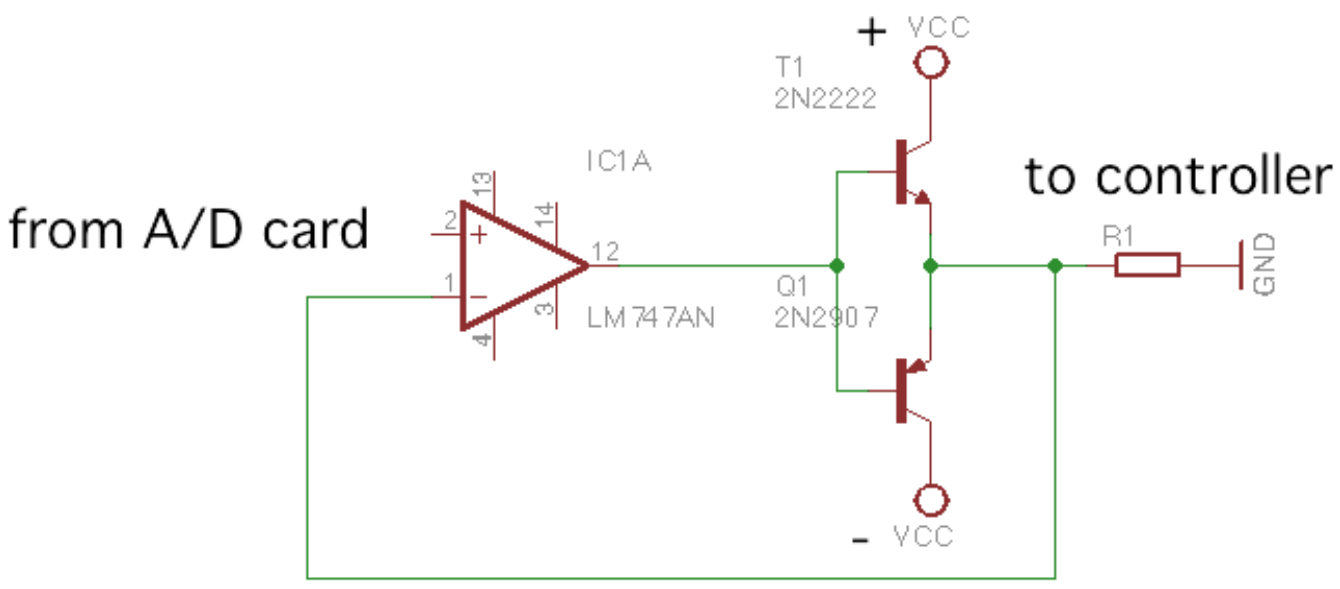

*Scheme 1*: electrical layout of the buffer circuit

The polar and azimuthal beam angles ($\theta, \phi$) were software controlled under LABVIEW (National Instruments, Austin, TX) using the amplified output of a 16-bit DAC board (NI U2B-6211, National Instruments) connected to the frequency modulation input of the VCO driver (AA.DRF.10Y2X, AA.Optoélectronique). We increased the output power of the DAC board by a custom buffer circuit with unity gain (scheme 1).

The excitation optical path was aligned using the BFP image taken on CCD camera '2', **Fig.S1***A*. For a perfectly symmetric illumination, the coverslip could finally be aligned orthogonally to the optical axis with a two-axes tip-tilt platform (M-TTN80, Newport, Irvine, CA).

**Set-up characterization.** Because no light is transmitted beyond the critical angle TIR limits the range within which the beam angles $\theta$ and $\phi$ can be calibrated. We therefore employed an oil-coupled solid-immersion half-ball lens (SIL, 8 mm ∅, S-LAH79, Edmund Optics, Barrington, NJ) that allowed us to project the beam beyond angles otherwise obscured by TIR (8, 9), *inset* in **Fig.S1***A*. At 488 nm, the S-LAH79 glass (Ohara, Branchburg, NJ) of the SIL has a refractive index of $n_3$ = 2.027 ± 0.002, refracting the beam to angles closer to the optical axis. The transmitted beam could then be projected onto a screen and the beam angle $\theta$ estimated and related to the measured radial and axial distances $a$ and $b$ as

$$\theta = \arcsin\left[n_3/n_2 \cdot \sin\left(\arctan(b/a)\right)\right], \quad \text{(eq.2)}$$

(and likewise for $\phi$ in orthogonal direction). Here, $n_2$ = 1.521 ± 0.006 is the refractive index at 488 nm of the BK-7 front lens of the TIRF objective. Multi-layer effects (10, 11) were neglected.

Phase aberrations were measured with a lateral shearing interferometric wave-front sensor (12) (SID4, Phasics, Palaiseau, France). $M^2$, the beam parameter product relative to that of an ideal Gaussian beam, and the Strehl ratio $S = e^{-(2\pi\sigma/\lambda)^2}$ ($\sigma$ being the RMS deviation of the wave front and $\lambda$ the wavelength) calculated and plotted as a function of $\theta$.

Due to the angle-dependent diffraction efficiency of the AODs the transmitted light intensity varied with spot position in the BFP, **Fig.S1***B*, and was further modified by the dichroic and objective, **Fig.S1***C*. These effects could, in principle, be automatically compensated for by feeding a correction signal into the intensity modulation input of the VCO, however, this possibility was not used here. Rather, we simply increased the laser power to have sufficient signal at high beam angles, where the diffraction efficiency of the AOD scanner is low. Typical laser powers in the sample plane were of the order of a few µW.

We used a 3-mm thick zt491 RDCXT (AHF, Tübingen, Germany) to direct the scanned 488-nm beam to the objective. Optical flatness ($\lambda$/10) and strain-free mount of the dichroic were crucial for optimal spTIRF. A small amount of the excitation light transmitted by the dichroic formed a scaled image ($f_1$ = 80 mm, $f_2$ = 40 mm, both from Linos) on an inexpensive CCD camera (Foculus FO432SB, Elvitec, Pertuis, France, 4.65 µm pixel size) permitting the simultaneous visualization of the excitation light distribution in the objective BFP (7) and of the fluorescence on a second, electron-multiplying charge-coupled device camera (EMCCD, Cascade 128+ or QuantEM512C, both from Photometrics, Tucson, AZ). Pixel sizes were 194 nm or 120 nm in the sample plane –calibrated with a G300-Cu copper grid (EMS, Hatfield, PA) or the 5-µm divisions of a reticle (Qioptiq-Linos), corresponding to an effective field-of-view of 25×25 µm² (61×61 µm²) for the Cascade 128+ (QuantEM512C). Combinations of the optical filters used are listed in **Table S5**. Filters were from AHF Analysentechnik (Tübingen, Germany), Barr Assoc. (Westford, MA), Chroma Technology (Bellows Falls, VT) and Semrock (Rochester, NY).

**Automatic thresholding, image segmentation and statistics.** EMCCD images were analyzed after subtraction of a dark image taken at the same exposure time, hardware- and amplifier-gain. Foculus CCD images are shown as raw images. Images are shown on an inverted look-up-table for better display in print with intensity displayed in grey value autoscaled. Circular line regions of interest (**Fig. 1***A*) were centered on the cell identified from bright-field (BF) images and the intensity profile plotted as a polar graph $I(\phi)/I_{mean}$ to allow cell-to-cell comparison, $CV = I_{SD}/I_{mean}$. Image segmentation and analysis (**Fig. 2***A*, **Fig. S2**) was performed using Metamorph's IMA tool (v7.5.4, Molecular Devices). To isolate individual lysosomes or mitochondria the local background was suppressed by subtraction of a corresponding low-pass filtered image (3.9 µm) and the result segmented using isodata histogram thresholding (13). Briefly, in this procedure the image histogram is initially segmented into two parts using a starting threshold intensity $T_0 = 2^{B-1}$, half of the maximum dynamic range. The sample mean ($m_{f,0}$) of the grey values associated with the foreground pixels and the sample mean ($m_{b,0}$) of the grey values associated with the background pixels are computed. A new threshold value $\theta_1$ is computed as the average of these two sample means, $(m_{f,0} + m_{b,0})/2$ . The process is repeated, based upon the new threshold, until $T_k$ converges,

$$T_k = (m_{f,k-1} + m_{b,k-1})/2 \;\; \text{until} \;\; T_k = T_{k-1} \quad \text{(eq.3)}$$

Individual mitochondria were identified using {area ≥ 5px (~ 1µm²) AND length ≥ 2px (0.38 µm) AND breadth ≥ 2px)} as classifiers and their area $A$ (pixels above $T_\infty$), orientation (angle between longest chord of the object and the horizontal, i.e., -90° = downward), shape factor ($\sigma = 4\pi A/P$, $P$ = perimeter, i.e., flat = 0, ..., 1 = circle), and elliptical form factor ($\varepsilon$ = length/breadth) measured. For the mitochondrial data set in **Fig. S2** (imaged at 78° beam angle and 194-nm pixel size), this corresponded to 14.4 ± 2.8% ($n$ = 24 images) of the maximal intensity. The means of normally and log-normally distributed data sets having the same variance were compared with Student's $t$-test. The non-parametric and distribution-free KS-test was used for comparing non-normally distributed data. Points ≥1.5× IQR above the third quartile or below the first quartile were considered as outliers (Turkey). Differences were considered significant for $p < 0.05$. On figures, *, **, and *** are shorthand for $p$ <

0.05, $p < 0.01$ and $p < 0.001$, respectively. *n.s.* means not significant.

## SUPPLEMENTARY TABLES

**Table S1:** *Measured microscope PSFs*

| | TIRF [a] | spTIRF | TIR-*v*SAF | spTIR-*v*SAF |
|---|---|---|---|---|
| **×100/1.46** | 366 ±19 [b] | 388 ± 19 [c] | 374 ± 20 | 375 ± 15 |
| **×60/1.45** | 336 ± 21 | 332 ± 25 | | |
| | 314 ± 26 [c] | 312 ± 14 | | |

[a] reported as FWHM (mean ± 1 SD) of a Gaussian fit (14) with the measured radial intensity profile of $n$ = 10 isolated in-focus red-fluorescent ( $\lambda_{ex}^{(max)}$ / $\lambda_{em}^{(max)}$ = 488/685 nm) 93-nm diameter beads.

[b] The PSF for the Zeiss ×100/1.46 lens is an overestimate because it was measured without the matched Zeiss tube lens that provides part of the aberration correction.

[c] same, for a second, nominally identical, PlanApo ×60/NA1.45oil TIRFM objective (Olympus).

**Table S2:** *Measured refractive indices of used immersion liquids*

| | | $n_D^{22-23}$ **(measured)** [a] | ***n*** **(nominal)** |
|---|---|---|---|
| **Olympus** | | 1.5122 ± 0.0009 | 1.516 |
| **Zeiss Immersol 518F** | (1) | 1.5156 ± 0.0002 | 1.518 |
| | (2) | 1.5157 ± 0.0002 | |
| | (3) | 1.5158 ± 0.0002 | |
| **Cargille FF** | | 1.4804 ± 0.0003 | 1.479 |

[a] measured at 22-23°C on a WYA Abbe refractometer using the sodium D line (589 nm). Values are mean ± 1 SD of five independent measurements for each immersion oil. Three different batches (1)-(3) of the Zeiss oil were tested.

**Table S3:** *Characteristics of polystyrene (PS) microsphere suspensions* [(a)]

| *d* (µm) | *x* [b] | *g* | *c* (µl$^{-1}$) | (µm$^{-3}$) | $\mu_s$ (mm$^{-1}$) | $\mu_s$' (mm$^{-1}$) |
|---|---|---|---|---|---|---|
| 0.7 | 6 | 0.912 | 0.1 | $10^{-10}$ | $10^{-10}$ | $8.8 \times 10^{-12}$ |
| | | | 10 | $10^{-8}$ | $9.5 \times 10^{-6}$ | $8.362 \times 10^{-7}$ |
| | | | 100 | $10^{-7}$ | $9.5 \times 10^{-5}$ | $8.362 \times 10^{-6}$ |
| | | | 1,500 | $1.5 \times 10^{-6}$ | $1.43 \times 10^{-3}$ | $1.258 \times 10^{-4}$ |
| | | | 15,000 | $1.5 \times 10^{-5}$ | 0.01425 | $1.258 \times 10^{-3}$ |
| | | | $1.5\times10^{5}$ | $1.5 \times 10^{-4}$ | 0.1425 | 0.0125 |
| | | | $1.5\times10^{6}$ | $1.5 \times 10^{-3}$ | 1.425 | 0.1254 |
| 2.8 | 20.6 | 0.813 | 0.1 | $10^{-10}$ | $10^{-10}$ | $1.87 \times 10^{-11}$ |
| | | | 10 | $10^{-8}$ | $1.58 \times 10^{-4}$ | $2.955 \times 10^{-5}$ |
| | | | 100 | $10^{-7}$ | $1.58 \times 10^{-3}$ | $2.955 \times 10^{-4}$ |
| | | | 1,500 | $1.5 \times 10^{-6}$ | 0.02362 | $4.416 \times 10^{-3}$ |
| | | | 15,000 | $1.5 \times 10^{-5}$ | 0.2362 | 0.0442 |
| | | | $1.5\times10^{5}$ | $1.5 \times 10^{-4}$ | 2.362 | 0.4416 |
| 90 | 579.4 | 0.929 | 0.01 | $10^{-11}$ | $10^{-10}$ | $7.1 \times 10^{-12}$ |
| | | | 0.6 | $6 \times 10^{-10}$ | 0.008 | $5.68 \times 10^{-4}$ |
| | | | 6 | $6 \times 10^{-9}$ | 0.078 | 0.0055 |
| | | | 60 | $6 \times 10^{-8}$ | 0.778 | 0.0552 |

[a] input parameters: $m_{PS} = n_{PS} - ik_{PS} = 1.6053 + 0$ (the imaginary part is negligible at visible wavelengths); $n_{medium} = 1.33$; $\lambda_0 = 488$ nm; $d$ – sphere diameter; $c$ – concentration in in µl$^{-1}$ (⇔ mm$^{-3}$)
[b] output parameters calculated with the Mie scattering calculator http://omlc.ogi.edu/calc/mie_calc.html

| | | |
|---|---|---|
| $x$ | - | size parameter $x = 2\pi d/\lambda$, where $\lambda = \lambda_0/n_{medium}$ |
| $g$ | - | average cosine of the phase function (scattering anisotropy) $g = \langle\cos\theta\rangle$ |
| $\mu_s$ | - | scattering coefficient |
| $\mu_s'$ | - | reduced scattering coefficient $\mu_s' = \mu_s \cdot (1-g)$ |

**Table S4:** *Plasmids transfected into astrocytes* [(a)]

| Plasmid | Concentration (µg/µl) | Source [(b)] |
|---|---|---|
| VAMP2-EGFP | 4.06 | Frank Kirchhoff (Univ. des Saarlands, Homburg; Germany) |
| YFP | 3 | *idem* |
| Mito-GFP | 0.25 | Johannes Hirrlinger (Univ. Leipzig, Germany) |
| Lck-EGFP | 2 | Steven Green (Univ. Iowa, USA) |
| CD63-GFP | 1 | Thierry Galli (Inst. Jacques Monod, Paris, France) |
| Vinculin-GFP | 2.4 | Maité Coppey (Inst. Jacques Monod, Paris, France) |

[a] using lipofectamine 2000 following standard protocols
[b] the kind gift of plasmids is gratefully acknowledged

**Table S5:** *Used filter combinations* [(a)]

| Fluorophore/experiment | dichroic | emission |
|---|---|---|
| FM2-10 | Q515LP | E580/40 |
| | zt491 RDCXT | E580/40 |
| FM4-64 | zt491 RDCXT | 600LP |
| EYFP/EGFP | Q515LP | 520LP |
| 488/560nm (488/685nm)TransFluoSpheres | zt491 RDCXT | BP530/40 (676/29BP) |
| Dark-field | zt491 RDCXT | HC488/10 or HC MaxLine 488/1.9 |

[a] 488-nm excitation was used in all experiments. Returned excitation light was stopped by a 488-nm rugate notch filter (Barr Assoc., Westford, MA) and z488rdc RazorEdge (Semrock, Rochester, NY) having a combined suppression of $>10^{-9}$.

# SUPPLEMENTARY FIGURES

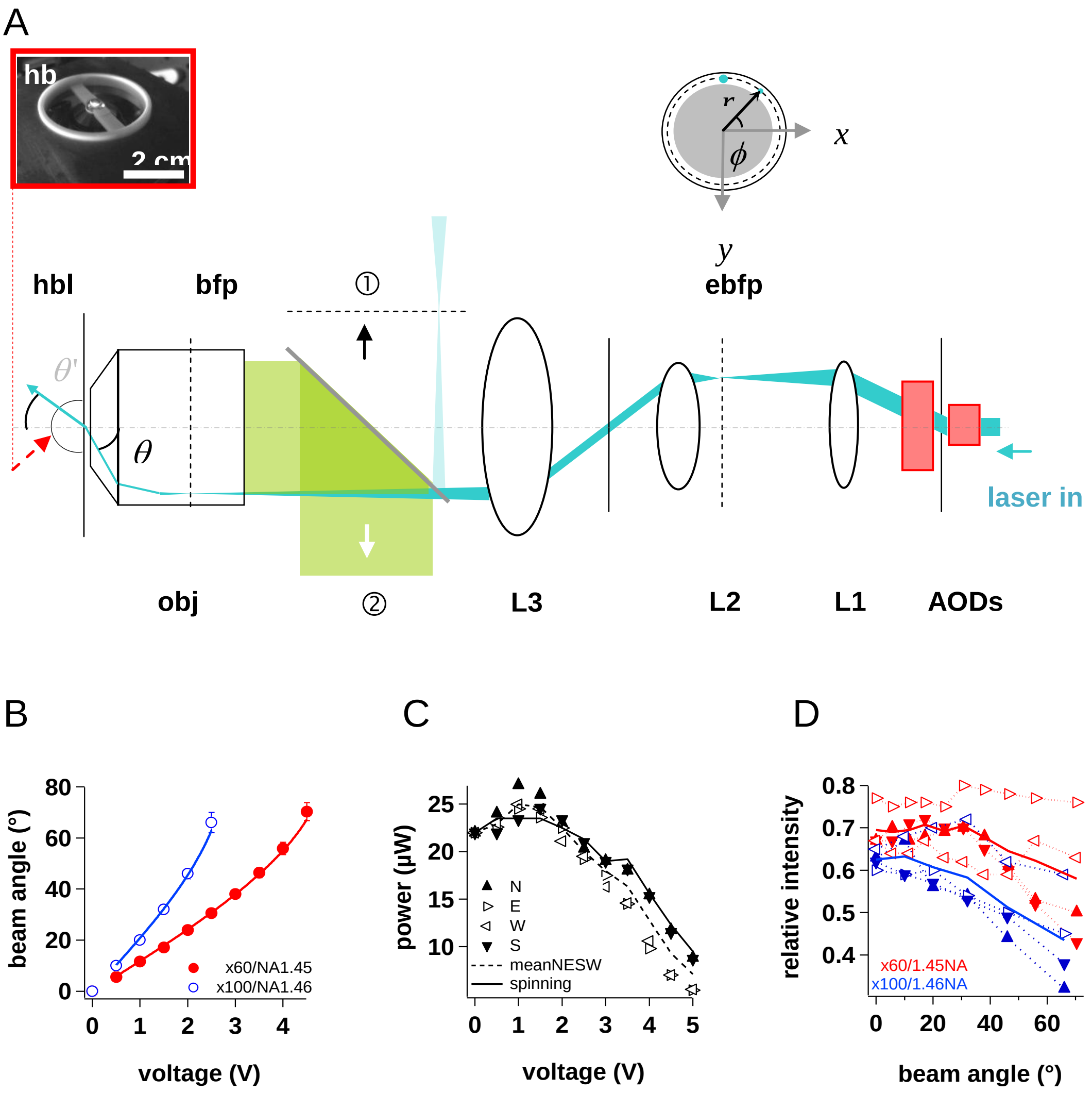

***Fig. S1***. *Spinning TIRF (spTIRF) excitation. (A), simplified optical layout of the excitation optical path. Solid and dashed lines are conjugate field and aperture planes, respectively. Optical elements are identified in the* ***SI materials*** *section. Insets show, on the left, photograph of the half ball lens (HBL) in a custom holder on top of the objective that allowed us to measure beam angles otherwise obscured by TIR. Right, definition of variables in the objective BFP. r and dashed line – radius; θ azimuthal angle; grey area – radii below $r_c$ corresponding subcritical angles at which light is refracted and propagated into the sample (epi); white annulus – radii corresponding to supercritical angles for which total internal reflection occurs; turquoise spot is focused excitation beam (not to size); solid line – radius $r_{NA}$ corresponding to limiting NA of the objective. (B), calibration of command voltage vs. beam angle* $\theta = \arcsin\left[\left(n_3/n_2\right)\sin\theta'\right]$ *for the ×60/NA1.45, red, and ×100/NA1.46 objective, blue.* $\theta'$ *is the measured angle of the beam exiting the HBL, $n_3$ = 2.03 its refractive index at 488 nm. Symbols and error bars represent center and diameter of the beam projected against a small screen. Fits with the objective transverse magnification M, $n_3$ and $f_{L3}$ as free parameters yielded (M, $n_3$, $f_{L3}$) = (58.83 ± 3.24, 2.04 ± 0.04, 137.82 ± 2.12, red) and (101.31 ± 2.2, 2.07 ± 0.05, 136 ± 3.56, blue), respectively, and compare favorably to the real values (60, 2.03, 135) and (100, 2.03, 135) for either objective. (C), measured laser power diffracted into order (1, 1) as a function of the command voltage of the AODs, for the four cardinal beam positions, symbols, their mean, dashed, and upon beam spinning, solid line. (D) Fractional intensity after the objective, as a function of beam angle θ for the four cardinal directions (symbols as in left panel) and their mean, solid. For all objectives, transmitted intensities were reduced at high NAs. At a given beam angle, laser power was adjusted to give the desired signal-to-noise ratio.*

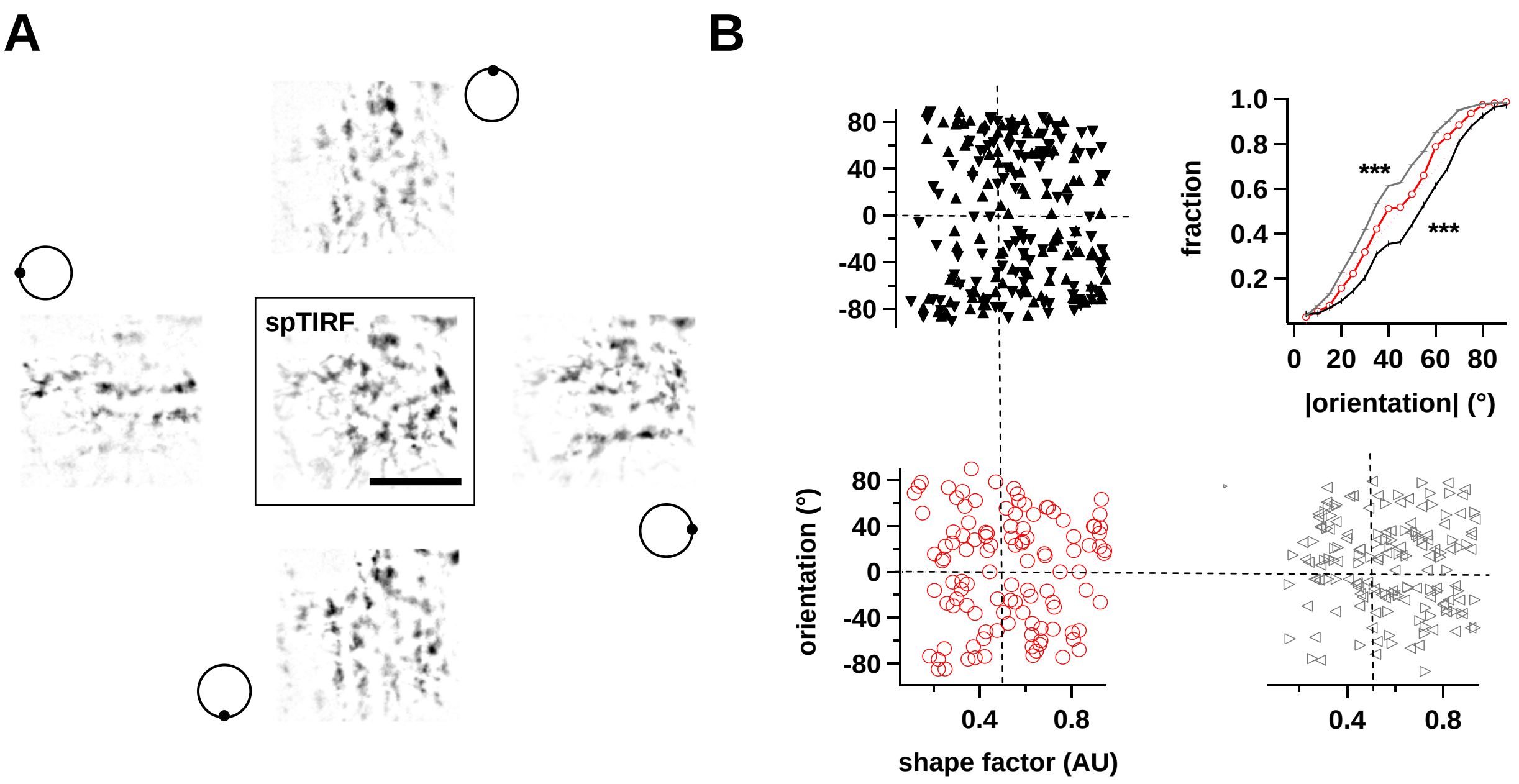

***FIG. S2.*** *Eccentric-spot excitation introduces a directional detection bias. (A), conventional unidirectional TIRFM images (cardinal images) of astrocytes expressing enhanced green fluorescent protein linked to a mitochondrial targeting sequence (mito-EGFP) show preferentially those mitochondria that are aligned parallel to the EW propagation direction, whereas those perpendicular to it are less visible. Symbols indicate position of the focused exciation spot in the objective BFP. (B), quantitative morphometry reveal that mitochondria detected upon illumination with a horizontally propagating EW (grey open triangles) had, on average, orientations closer to the horizontal axis, (33.7° ± 16.6°, median ± abs. deviation, n = 268 mitochondria in 6 cells), whereas those detected with a 90°-rotated EW field (black filled triangles) were orientated closer to the vertical axis (53.2° ± 18.6°, n = 357, p < 0.001). Triangles point in direction of EW propagation. Restoring a symmetric illumination by azimuthal beam scanning abolished this directional detection bias (red circles), and the cumulative distribution of absolute orientations (between 0 and 90°) now had a median at the expected 45° (p < 0.001 vs. each of the four unidirectional images, n = 155 mitochondria in 6 cells). Color code as on panels of orientation vs. shape factor that is 4πA/P², where A and P are the organelle area and perimeter, respectively.*

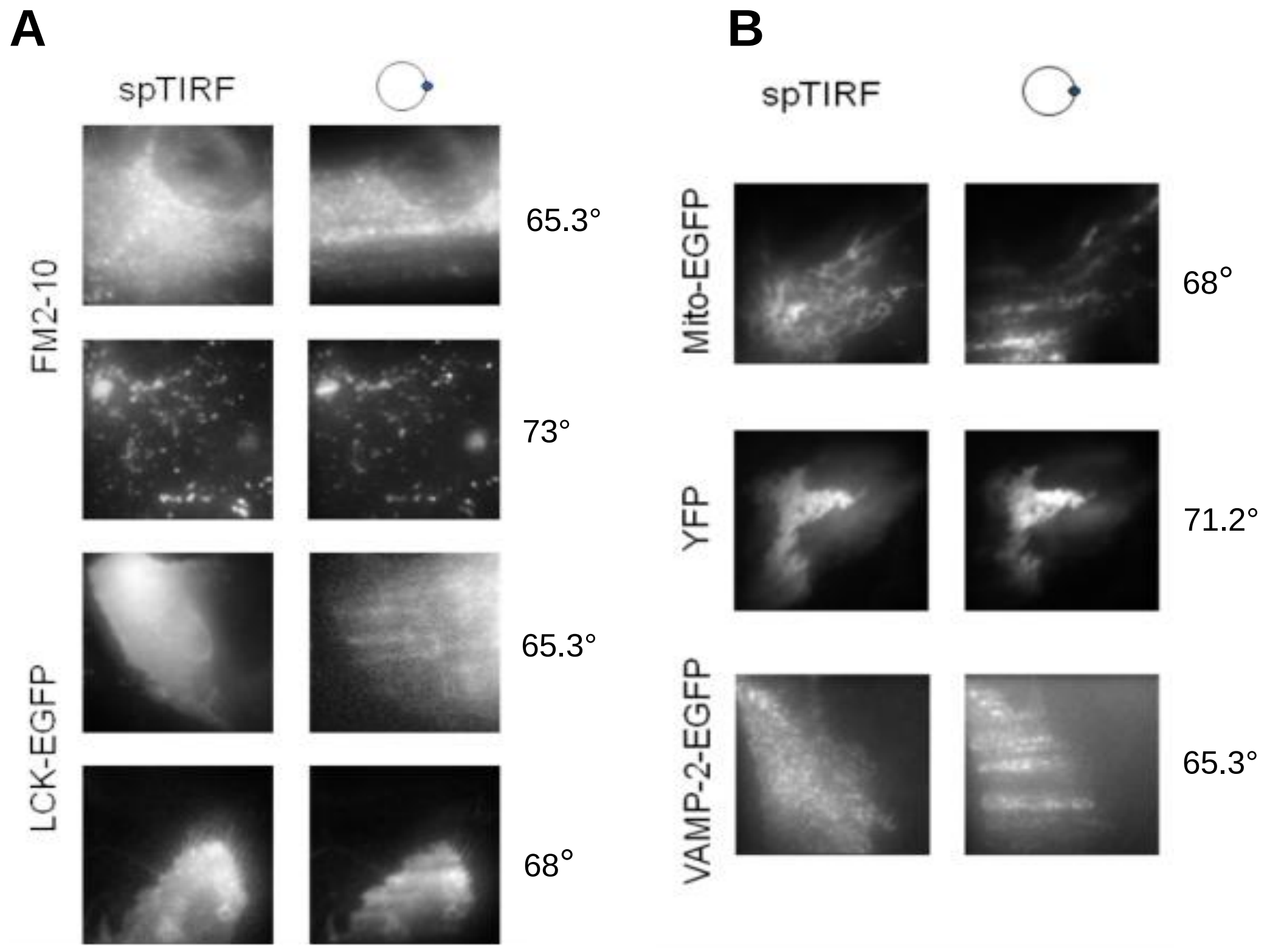

***Fig. S3.*** *High fluorophore density and large penetration depth aggravate excitation non-homogeneity. Image pairs show cultured cortical astrocytes, (A), labeled with FM2-10, after transfection with various fluorescent-protein constructs, upon azimuthal beam spinning (left, spTIRF) and unidirectional 488-nm evanescent-wave excitation (right row). Denser fluorophore distributions like membrane (Lck-EGFP), cytoplasmic (YFP) or dense vesicular labeling (vesicular-membrane associated protein-2, VAMP-2) are more sensitive reporters of non-evanescent excitation light than are sparser labels like lysosomes (FM2-10). Uneven illumination is less perceptible at higher beam angles (smaller penetration depths of the evanescent wave). In all cases, azimuthal beam spinning effectively eliminates excitation patterns. Objective was ×60/NA1.45. Fields are 25×25 µm², pixel size 194 nm in the sample plane. See* ***table S4*** *for used filter combinations.*

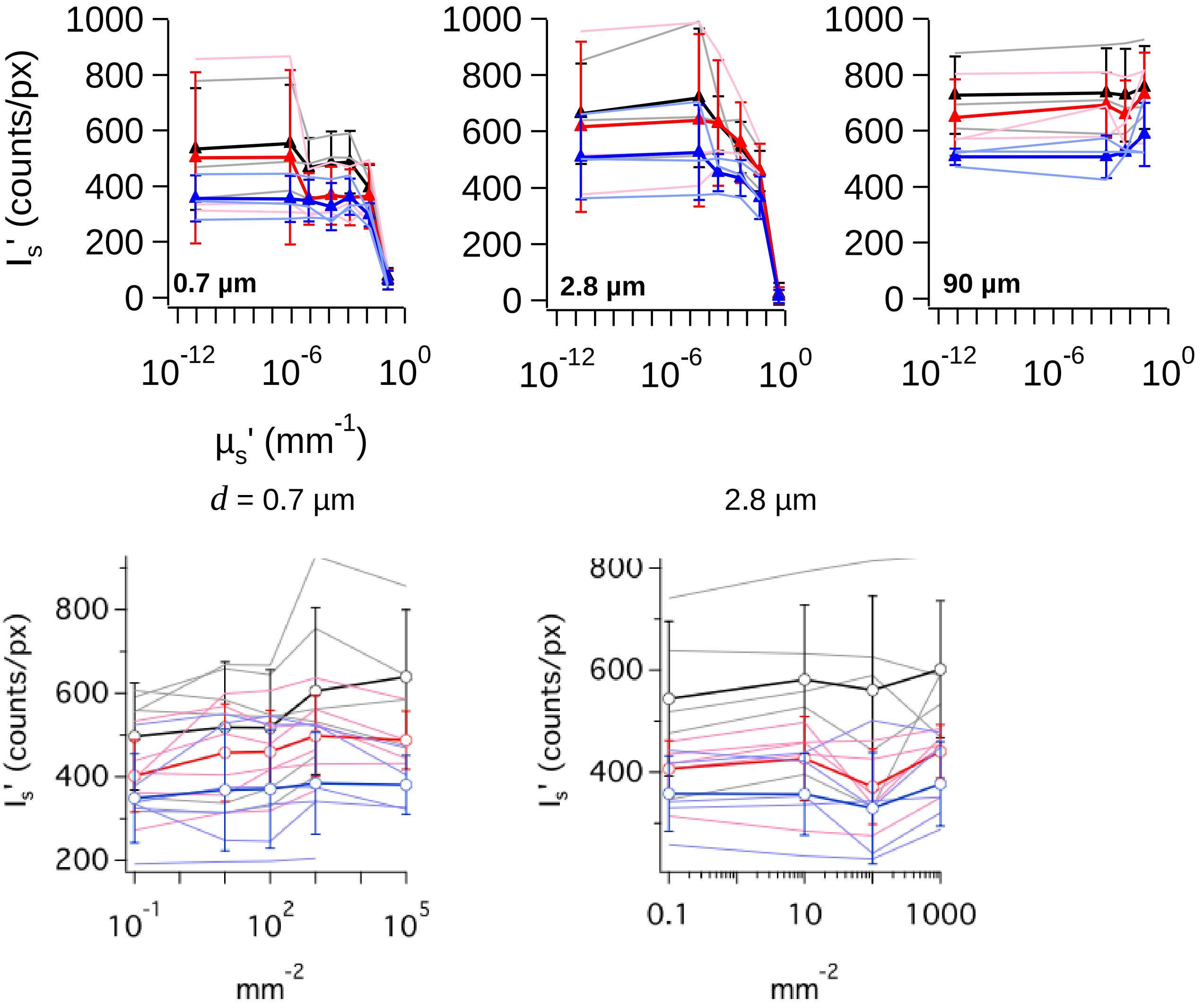

***FIG. S4**. Effect of evanescent-field scattering. (A), dependence of the scattered light intensity $I_s$' on the reduced scattering coefficient $\mu_s$' of dilute solutions of non-fluorescent polystyrene beads, for three different beam angles $\theta$ (black, 64°; red, 70° and blue, 78°) and three different bead diameters, from left to right: 0.7, 2.8 and 90 µm. For each color, solid lines and symbols show mean ± SD from triplicate experiments, thin faint lines are individual measurements. Drop at high values of $\mu_s$' is due to inner-filter effects due to multiple scattering. Scattered light was detected in dark-field through a second upright ×60/1.1 water-immersion objective the focus of which was at the reflecting interface (see **Fig. 3**A). The EW was set up by focusing a 488-nm laser beam in the periphery of the lower ×60/1.45 oil-immersion objective. (B), Dependence of $I_s$' on the surface density of scattering beads (mm$^{-2}$), for a monolayer of 0.7- (left) and 2.8-µm beads (right) drop-cast on the coverslip. Color code as before. Compare with **Fig. 3B** in the main text.*

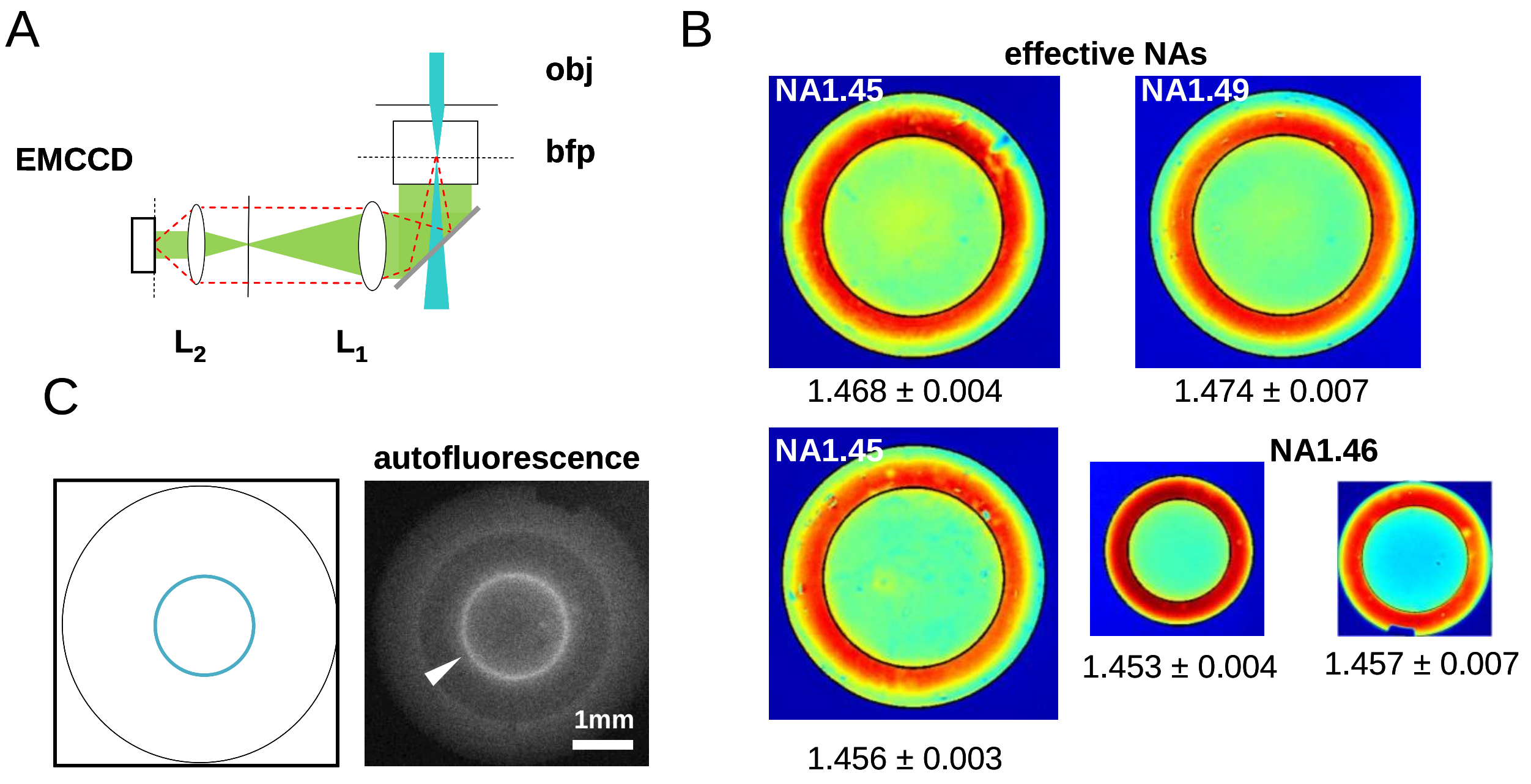

***Fig. S5**. Properties of high-NA objectives used. (A), schematic optical layout. $L_1$ – AC250 mm (Linos), $L_2$ – AC110 mm (TILL). EMCCD – QuantEM512C. (B), measured intensity distribution at the BFP for, from left to right, two nominally identical Olympus ×60/NA1.45 objectives, the Olympus ×60/NA1.49 and the Zeiss ×100/NA1.46 objective, $r_c$ and $r_{NA}$ indicated by black solid lines are, respectively, the radii corresponding to the critical and maximal emission angle supported by the objective and are related to $\theta_{NA}$ by $r = f \sin(\theta_{NA})$, where $f = f_{TL}/M$ and $\theta_{NA} = \arcsin(n_2/NA)$ are the objective focal length and aperture angle, respectively. Sample was a 100-nm thick FITC layer (500 µM). Laser powers in the back pupil were 226, 104, 117 and 100 µW, respectively. Measured effective NAs ($NA_{eff}$) are shown below the corresponding images, the error giving the SD of independent measurements (re-aligning the Bertrand lens, re-focusing the objective or moving laterally in the fluorophore layer). (C), (left) sketch of BFP upon focusing a low-µW 488-nm beam in the BFP of a Zeiss x100/NA1.46 objective. The beam was spun at constant a radius, producing in several lenses measureable yellow-green autofluorescence (right image), the most intense of which originating from glass close to the BFP (arrowhead).*

## SUPPORTING REFERENCES